\documentclass[aps,twocolumn,superscriptaddress]{revtex4-1}
\usepackage[utf8]{inputenc} 
\usepackage[T1]{fontenc}    %
\usepackage[english]{babel} 
\usepackage{amssymb,amsmath,mathtools} 
\usepackage{graphicx} 
\usepackage{xcolor}

\usepackage{hyperref}

\newcommand{\Tr}{{\rm Tr}\,}

\begin{document}

\title{Spacing distribution in the 2D Coulomb gas: Surmise and symmetry classes of non-Hermitian random matrices at non-integer $\beta$}

\author{Gernot Akemann}\email{akemann@physik.uni-bielefeld.de}\affiliation{Faculty of Physics, Bielefeld University, Postfach 100131, 33501 Bielefeld, Germany}
\author{Adam Mielke}\email{admi@dtu.dk}\affiliation{Technical University of Denmark, 
Asmussens All\'e, Building 303B, 2800 Kgs. Lyngby, Denmark}
\author{Patricia P\"a{\ss}ler}\email{patricia@physik.uni-bielefeld.de}\affiliation{Faculty of Physics, Bielefeld University, Postfach 100131, 33501 Bielefeld, Germany}

\date{\today}

\begin{abstract}

A random matrix representation is proposed for the two-dimensional (2D) Coulomb gas at inverse temperature $\beta$. For $2\times 2$ matrices with Gaussian distribution 
we analytically compute the nearest neighbour spacing distribution of complex eigenvalues in radial distance. Because it does not provide such a good approximation as the Wigner surmise in 1D, we introduce an effective $\beta_{\rm eff}(\beta)$ in our analytic formula, that describes the spacing obtained numerically from the 2D Coulomb gas well for small values of $\beta$. It reproduces the 2D Poisson distribution at $\beta=0$ exactly, that is valid for a large particle number. The surmise is used to fit data in two examples, from open quantum spin chains and ecology. The spacing distributions of complex symmetric and complex quaternion self-dual ensembles of non-Hermitian random matrices, that are only known numerically,  are very well fitted by {\it non-integer values} $\beta=1.4$ and $\beta=2.6$ from a 2D Coulomb gas, respectively. These two ensembles have been suggested as the only two symmetry classes, where the 2D bulk statistics is different from the Ginibre ensemble. 

\end{abstract}
 
\date{\today}
\maketitle

\section{Introduction}
The description of the spacing between neighbouring  energy levels by the distribution following from the three Wigner-Dyson classes of random matrices is probably the most used quantity in this field. A simple, approximate formula exists, the Wigner surmise,  that following the Bohigas-Giannoni-Schmit conjecture \cite{BGS,CGV} applies to fully chaotic quantum Hamiltonians. The symmetry class depends on the presence or absence of time-reversal and spin taking integer or half-integer values, Dyson's threefold way \cite{Dyson}. 
The joint density of eigenvalues of these three classical ensembles of random matrices equals a 2D Coulomb gas confined to the real line at three different values of the inverse temperature $\beta=1,2,4$ in a Gaussian potential, the so-called Dyson gas. 
Many applications in Physics beyond quantum chaos exist \cite{GMW,ABD}, and the predictions are know to hold beyond the Gaussian potential \cite[Chap. 6, 9]{ABD}, thus being universal.

The idea to describe many-body chaotic quantum systems using Hermitian random matrices goes back to the concept of the compound nucleus by Bohr, see \cite{GMW}.
While the consideration of non-Hermitian random matrices with complex spectra was initially born out of mathematical curiosity \cite{Ginibre}, nowadays there are many applications to non-Hermitian operators with a two-dimensional (2D) spectrum. Examples include localisation 
in 2D random Schr\"odinger operators \cite{HatanoNelson}, dissipative quantum systems \cite{GHS}, random neural networks \cite{SCS},  
quantum field theory with a quark chemical potential \cite{Wettig}, 
the 3D Anderson model with disorder \cite{HS}, 
and beyond physics in the adjacency matrix of directed complex networks \cite{YQWG} 
or ecology \cite{ABC}. In particular, open chaotic quantum system have seen much activity recently, where the effect of integrability vs. chaos is studied in spin chains \cite{AKMP,HKKU,RGMD} or the kicked rotor \cite{JPP}. Apart from these applications, the 2D Coulomb gas (2DCG) is fascinating in its own right as a statistical mechanics problem, cf. \cite{Baus}, displaying the phenomenon of crystallisation \cite{AJ,CC}, see \cite{Abanov} for recent numerics and \cite{Sylvia} for a mathematical physics perspective.

As in 1D, the simplest null models proposed to describe spectra in 2D are Poisson random variables for completely uncorrelated points,  and complex eigenvalues of non-Hermitian random matrices for strong correlations, based on  symmetry. 
The conjectured correspondence to quantum integrable, respectively chaotic systems has been extended to 2D \cite{GHS}. 
In particular, it was argued based on perturbation theory \cite{GHS,Haake} that the nearest neighbour spacing in radial distance in 2D is universal at small distance and displays a cubic level repulsion, irrespectively of time-reversal being preserved or not \cite{GH}.

Perhaps surprisingly, the spacing distribution in the three non-Hermitian Ginibre ensembles all agree in the bulk of the spectrum away from the real axis \cite{GHS,BS,AKMP}. This is in stark contrast to the three Hermitian Wigner-Dyson classes where the matrices with real, complex or quaternion elements display a different level repulsion, governed by the respective inverse temperature $\beta=1,2,4$ of the Dyson gas.
Is there more than a single symmetry class in the bulk in 2D and is there a simple formula describing these?
In 1D, Dyson's classification \cite{Dyson} was extended to 10 classes by Altland and Zirnbauer \cite{AZ}. However, this did not extend the 3 classes of local bulk statistics, as additional symmetries only manifest themselves at specific points of the spectrum. Likewise, based on the symmetry classes of Dirac operators \cite{BLeC}, a first classification of non-Hermitian random matrices using symmetric spaces was undertaken in \cite{Magnea}, and was recently revisited \cite{KSUS}. It leads to 38 classes, cf. the most recent arXiv version v2 of \cite{BLeC}. 
Half of these classes have been identified in a non-Hermitian Sachdev-Ye-Kitaev model \cite{GSV}.
Based on heuristic arguments and numerics, it was found that 
only three distinct classes of 2D bulk statistics exist \cite{HKKU}: The Ginibre ensemble \cite{Ginibre} and two further classes labelled AI$^\dag$ and AII$^\dag$  (in analogy to \cite{AZ}), 
that possess additional symmetries under transposition. 
They are respectively given by  complex symmetric (AI$^\dag$) and complex quaternion self-dual matrices (AII$^\dag$) \cite{HKKU} (related to complex antisymmetric matrices \cite{Hastings}). 
The spectral statistics of AI$^\dag$  was found in the complex spectrum of the kicked rotor \cite{JPP}, where a transition from time-reversal invariance in AI$^\dag$ 
and its breaking towards the Ginibre ensemble was observed. 
A phase transition between class AI$^\dag$ and 2D Poisson was detected in \cite{HS}.
All three classes were found in the spectrum of Liouville operators of dissipative quantum spin chains, subject to certain external fields \cite{HKKU}, and in non-Hermitian Dirac operators in quantum field theories \cite{TT}. 
For questions of dynamics and non-equilibrium of Liouville operators, and a comparison of their global statistics to non-Hermitian random matrices, see  
\cite{Karol,Can1,Can2,TP2}.

It is the goal of this article to find an approximate description of the spacing distribution of these 3 classes (Ginibre, AI$^\dag$, AII$^\dag$) in terms of a 2DCG at inverse temperature $\beta$. Furthermore, a simple formula or surmise will allow us to describe the transition to 2D Poisson at $\beta=0$. This is in contrast to 1D, where Wigner's surmise 
\begin{eqnarray}
\label{pWigner}
&&p_{\rm Wigner}^{\rm 1D}(s)=a_\beta s^\beta \exp[-b_\beta s^2], \\
&&b_\beta=\Gamma[(\beta+2)/2]^2/\Gamma[(\beta+1)/2]^2,\ \ a_\beta=2b_\beta^{(\beta+1)/2}, 
\nonumber
\end{eqnarray}
works very well at increasing values of  $\beta=1,2,4$, 
but fails closer to Poisson in 1D,  
\begin{equation}
\label{Poisson1D}
p_{\rm Poisson}^{\rm 1D}(s)=\exp[-s],
\end{equation}
corresponding to $\beta=0$. Several phenomenological  interpolations between $\beta=0$ and 1 exist, cf. \cite{GMW}.
More recently, a $2\times2$ invariant matrix representation of the 1D Dyson gas for $\beta\in[0,2]$ has been used to derive an integral representation \cite{PierSatya}. We will follow this approach and extend it to a 2DCG at $\beta\geq0$, also called non-Hermitian $\beta$-ensemble, staying well below the crystallisation transition at 
$\beta_c\approx 140$ \cite{Abanov}. 

The local statistics in $\beta$-ensembles is a very active field in mathematical physics in 1D and 2D, see \cite{Sylvia} for a review on 2D. Even in 1D in the bulk, where there is a limiting point process in terms of a stochastic operator, the sine-$\beta$  process \cite{VV}, no explicit expression is known for the spacing distribution for $\beta\neq 1,2,4$.
In 2D closed form expressions for the local statistics exist only for the spacing for Poisson ($\beta=0$), see \eqref{Poisson2D} below, and the complex Ginibre ensemble (GinUE) at $\beta=2$, see \eqref{p2Gin}. Consequently, a detailed comparison to the spacing in the intermediate regime previously relied on numerical simulations of the 2DCG, cf. \cite{AKMP}. 
When comparing to data, also in 2D spectra have to be unfolded which is quite nontrivial, see \cite{Haake,Wettig,AKMP}. For that reason, complex valued ratios between nearest and next-to-nearest neighbour spacings have been proposed as an alternative  measure \cite{SRP}.

The remainder of this short article is organised as follows. In Section \ref{Invariant-beta} we construct an invariant non-Hermitian $\beta$-ensemble for complex normal matrices $J$, first for general size $N$. In Subsection \ref{Invariant-beta-N2}, $J$ is parametrised for $N=2$, leading to an explicit expression for the spacing distribution and spectral density for general $\beta\geq0$. 
In Section \ref{Surmise} we approximate the spacing distribution for general $\beta$ at large-$N$ which is generated numerically. The spacing for $N=2$ from Section \ref{Invariant-beta} is used and improved by introducing an effective $\beta_{\rm eff}$, 
by fitting $\beta$ to the numerically generated 2DCG at $N=5000$.  
This approximation is illustrated upon data with small to intermediate values of $\beta$. In Section \ref{SymmClass} we approximate the numerically generated spacing distribution of symmetry classes AI$^\dag$ and AII$^\dag$ by that of a non-Hermitian $\beta$-ensemble, finding non-integer values for the fitted $\beta$s.

\section{Invariant Non-Hermitian $\beta$-Ensemble}\label{Invariant-beta}
\subsection{Construction for general N}\label{Invariant-beta-genN}

Let us 
define the joint density of points (charges) 
of the 2DCG at fixed inverse temperature $\beta>0$, subject to a confining potential $V$, as
\begin{equation}
\label{2DCoulomb}
\mathcal{P}_{\beta,N}(z_1,\ldots,z_N) \propto
e^{\beta\sum_{i> j}^N\log|z_i-z_j|
-{\sum}_{l=1}^N V(z_l,z_l^*)
}.
\end{equation}
In order to be able to take the limit $\beta\to0$ of uncorrelated points, cf. \cite{ABy}, we have rescaled the potential to remove its dependence on $\beta$. An explicit random matrix representation is only known for $\beta=2$, given by  the complex eigenvalue distribution of the GinUE \cite{Ginibre} with Gaussian potential $V(z,z^*)=|z|^2$.
Assuming that the bulk of the spectrum is translation invariant, we can compute the nearest neighbour spacing distribution in 2D by putting a point at the origin, and then determine the probability to find the next point at radial distance $s$,
\begin{equation}
\label{pbeta}
p_{\beta,N}(s) \propto 
\prod_{j-2}^N\int
_{\mathbb{C}}
\mathrm{d}^2z_j\mathcal{P}_{\beta,N}(0,z_2,\ldots,z_N)\,  \delta(|z_2|-s).
\end{equation}
For $\beta=2$ this expression can be computed analytically as given in  \eqref{p2Gin}, see \cite{GHS,Mehta} for a derivation.

Let us construct an ensemble of $N\times N$ complex normal non-Hermitian random matrices $J\neq J^\dag$, with $[J,J^\dag]=0$, that has the same joint distribution of complex eigenvalues as the 2DCG \eqref{2DCoulomb}, for arbitrary $\beta>0$. Here $\dag$ denotes the Hermitian conjugate, $J^\dag=J^{*T}$. 
Following \cite[Chap. 15]{Mehta} the Vandermonde determinant 
\begin{equation}
\label{Vandermonde}
\Delta_N(Z)= \prod_{N\geq i> j\geq 1}(z_i-z_j) 
\end{equation}
of complex eigenvalues of $J$, that provides the logarithmic Coulomb interaction in \eqref{2DCoulomb}, can be written as follows, after taking the modulus squared,
\begin{eqnarray}
\label{Delta-rewrite}
&&|\Delta_N(Z)|^2=\left|\det \begin{pmatrix}
	1&...&1\\
	z_1&...&z_N\\
	...&...&...\\
	z_1^{N-1}&...&z_N^{N-1}
\end{pmatrix} \right|^2
\nonumber \\
&&=\det \begin{pmatrix}
N& \sum_iz_i&...&\sum_iz_i^{N-1}\\
\sum_i{z}^*_i& \sum_i{z}^*_iz_i&...&\sum_i {z}^*_iz_i^{N-1}\\
...&...&...\\
\sum_i{z}_i^{*N-1}& \sum_i{z}_i^{*N-1}z_i&...&\sum_i {z}_i^{*N-1}z_i^{N-1}
\end{pmatrix} \nonumber \\
&&= \det_{1\leq i,j\leq N}\left[\Tr(J^{i-1}J^{\dag\,j-1})\right].
\end{eqnarray}
Here, we used $|\det[A]|^2=\det[A]\det[A]^*$ in the first line and multiplied these two determinants row by row, as one of them can be replaced by its transpose. In the third step we used the that the normal matrix $J$ can be diagonalised via a unitary transformation, i.e. $J=UZU^\dag$ with $U\in U(N)$ and $Z=$diag$(z_1,\ldots,z_N)$, which yields 
\begin{equation}
\Tr (J^kJ^{\dag\,l})=\sum_{i=1}^Nz_i^k{z}^{*l}_i, \quad k,l=0,1,...,N-1.
\end{equation}
 We obtained the modulus square of the Vandermonde determinant expressed in rotational invariant terms.
Because the Jacobian for the above stated diagonalisation of complex normal matrices $J$  is known \cite{Oleg} to be proportional to  
$|\Delta_N(Z)|^2$, we obtain the following distribution
\begin{eqnarray}
\mathcal{P}_{\beta,N}(J)&\propto& \frac{\exp[-\Tr V(J,J^\dag)]}{\det_{1\leq i,j\leq N}\left[\Tr(J^{i-1}J^{\dag\,j-1})\right]^\eta}
\nonumber\\
&\propto& |\Delta_N(Z)|^{2-2\eta}
e^{-{\sum}_{l=1}^N V(z_l,z_l^*)},
\label{betaRM}
\end{eqnarray}
for a non-Hermitian $\beta$-ensemble. 
It agrees with \eqref{2DCoulomb} when identifying $\beta=2-2\eta$. 
The integrals converge for $\beta\geq0$ or $\eta\leq1$ (however, cf. \cite{ABy} when $\beta=2c/N$ for $c> -2$).
This allows us to interpolate between the Poisson point process at $\beta=0$ and the GinUE at $\beta=2$, or larger values of $\beta$, when making $\eta<0$. Although this $\beta$-ensemble \eqref{betaRM} is invariant and well defined for any $N$, we have so far only been able to obtain analytical results for $N=2$ with  Gaussian potential, following \cite{PierSatya} closely.

\subsection{Analytic results for $N=2$: Parametrisation, density and spacing distribution}\label{Invariant-beta-N2}
We now study the model \eqref{betaRM} in detail for $N=2$ for a Gaussian potential $V(J,J^\dag)=JJ^\dag$,
\begin{eqnarray}
	\mathcal{P}_{\beta,N=2}(J)&\propto&
	\frac{\exp[-\Tr(JJ^\dag)]}{\left[2\Tr(JJ^\dag)-
		\Tr( J)\Tr( J^\dag)\right]^\eta}.
	\label{betaRMN2}
\end{eqnarray} 
It is not difficult to show that any complex normal $2\times 2$ matrix can be parametrised by the following 6 real parameters, $a_1,a_2,b_1,b_2\in\mathbb{R}$, $k_1\in[0,2\pi)$ and $k_2\in[0,\infty)$:
\begin{equation}
	J=\left( \begin{matrix}
		a & b\\
		b^*e^{2ik_1} & a+k_2 e^{ik_1} \\
	\end{matrix} \right),\quad a=a_1+ia_2,\ b=b_1+ib_2.
	\label{parametrization}
\end{equation}
This follows by imposing the condition $[J,J^\dag]=0$ onto a complex matrix $J$, and using polar coordinates for certain matrix elements. 
On the other hand, any complex normal matrix can be diagonalised by a unitary transformation $U$ \cite{GL}, with $U\in U(2)/U(1)^2$ in our case $N=2$. This can be parametrised as follows, see e.g. \cite{Haake}:
\begin{equation}
	U=\left( \begin{matrix}
		\cos\theta & -e^{-i\phi}\sin\theta\\
		e^{i\phi}\sin\theta & \cos \theta\\
	\end{matrix} \right),\quad \theta,\phi\in[0,2\pi).
\end{equation}
Thus the diagonalisation reads 
\begin{equation}
	\label{diag}
	J=\left( \begin{matrix}
		a & b\\
		b^*e^{2ik_1} & a+k_2 e^{ik_1} \\
	\end{matrix} \right)
	=U^\dag\left( \begin{matrix}
		z_1 & 0\\
		0 & z_2\\
	\end{matrix} \right) U, 
\end{equation}
including the complex eigenvalues $z_1,z_2\in\mathbb{C}$. As a consistency check this leads to the Jacobian \cite{P}
\begin{equation}
	\left| \det\left(\frac{\partial (a,b,k_1,k_2}{\partial (z_1,z_2,\phi,\theta)}\right)\right|\sim|z_2-z_1|^2,
\end{equation}
for the change of variables \eqref{diag}. Because the Jacobian from the diagonalisation of $J$ is known to give $|z_2-z_1|^2$ for $N=2$, or the  absolute value squared of the Vandermonde determinant for general $N$ \cite{Oleg}, the parametrisation \eqref{parametrization} can only lead to a constant Jacobian.
Together with \eqref{Delta-rewrite} for $N=2$
\begin{equation}
	2\Tr(JJ^\dag)-\Tr(J)\Tr(J^\dag)=|z_2-z_1|^2,
\end{equation}
we indeed obtain from \eqref{betaRMN2} the normalised joint distribution of complex eigenvalues 
\begin{eqnarray}
	\mathcal{P}_{\beta,2}(z_1,z_2)&=&K_\beta|z_2-z_1|^{\beta}
	\exp[-|z_1|^2-|z_2|^2],
	\label{betaZN2}\\
	K_\beta^{-1}&=&\pi^22^{\beta/2}\Gamma[1+\beta/2],\nonumber
\end{eqnarray}
an $N=2$ $\beta$-ensemble at $\eta=1-\beta/2$. The normalisation constant $K_\beta$ is derived in Appendix \ref{jpdf}.\\

\begin{figure}[t!]
	\centering
	\includegraphics[width=0.49\linewidth,angle=0]{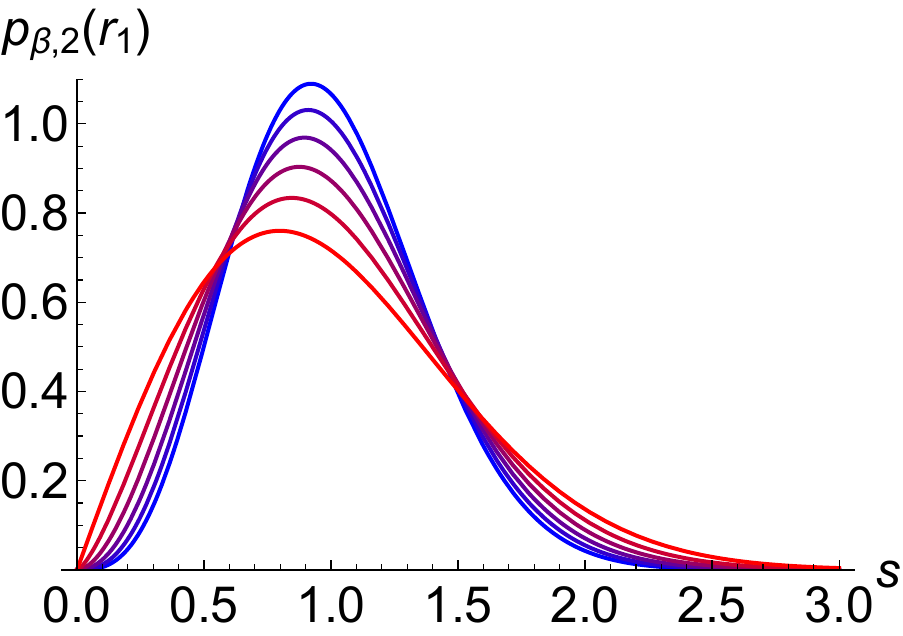}
	\includegraphics[width=0.49\linewidth,angle=0]{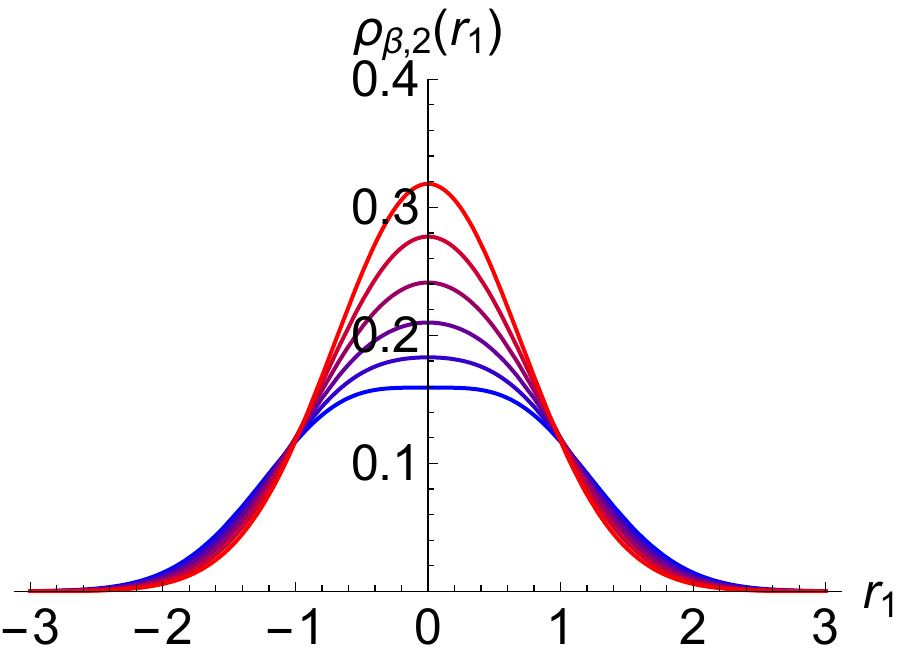}
	\caption{Left: spacing distribution $p_{\beta,2}(s)$ \eqref{peta} at $N=2$ for several values of $\beta=0,0.4,0.8,1.2,1.6$ and 2, where the blue top curve corresponds to $\beta=2$ (GinUE) and the red bottom curve to $\beta=0$ (2D Poisson). Right:  spectral density $\rho_{\beta,2}(r_1)$ at $N=2$ for the same values of $\beta$ from \eqref{average density model}. Here the red top curve represents Poisson and the blue bottom curve the complex Ginibre ensemble.
	}
	\label{Fig:spacing average density}
\end{figure}

Next we derive the spectral density and the spacing distribution for $N=2$. The spectral density  $\rho_{\beta,2}(z_1)$ is defined as  
\begin{eqnarray}
\rho_{\beta,2}(z_1)&=&\int_\mathbb{C}\mathrm{d}^2z_2\mathcal{P}_{\beta,2}(z_1,z_2)\nonumber\\
&=& K_\beta\int_\mathbb{C}\mathrm{d}^2z_2|z_2-z_1|^{\beta} e^{-|z_1|^2-|z_2|^2},
\label{rhodef}
\end{eqnarray}
and  normalised to unity, $\int_\mathbb{C}\mathrm{d}^2z_1\rho_{\beta,2}(z_1)=1$. 
We can follow the same lines as in the determination of the normalisation constant $K_\beta$ in Appendix \ref{jpdf}, by changing variables $z_2\to z=z_2-z_1=r_2e^{i\Theta_2}$, cf. \eqref{Kstep1}. Performing the angular integrations,  
the result is rotationally invariant, only depending on $|z_1|=r_1$:
\begin{eqnarray}
	\label{average density model}
	\rho_{\beta,2}(z_1)&=& K_\beta 2\pi e^{-2r_1^2}\int_{0}^\infty \mathrm{d}r_2 r_2^{1+\beta} e^{-r_2^2} I_0(2r_1r_2)
	\nonumber\\
	&=& \frac{1}{2^{\beta/2}\pi}\exp[-2r_1^2]\ {}_1F_1(1+\beta/2;1;r_1^2).
\end{eqnarray}
Here, ${}_1F_1(a;b;x)$ is Kummer's hypergeometric function, and in the last step we have applied \cite[6.631(1)]{Gradshteyn}, inserted $K_\beta$ from \eqref{betaZN2} and cancelled common factors. At the special values $\beta=2$ (GinUE), respectively $\beta=0$ (Poisson) we obtain for the density 
at $N=2$ 
\begin{align}
	\rho_{\beta=2,2}(r_1)&=\frac{1}{2\pi}(1+r_1^2)\exp[-r_1^2]\quad \mbox{(GinUE)},\\
	\rho_{\beta=0,2}(r_1)&=\frac{1}{\pi}\exp[-r_1^2] \quad\quad\quad \mbox{(Poisson)}.
\end{align}
This follows from the Taylor series
\begin{equation}
	{}_1F_1(a;b;x)=1+\frac{a}{b}x+\frac{a(a+1)}{b(b+1)2!}x^2+\ldots,
\end{equation}
and the connection formula \cite[13.2.39]{NIST}
\begin{equation}
	{}_1F_1(a;b;x)=e^x{}_1F_1(b-a;b;-x),
\end{equation}
at $a=2, b=1$ $(\beta=2)$ and $a=b=1$ ($\beta=0$), respectively.
The right diagram in Figure \ref{Fig:spacing average density} displays the spectral density  \eqref{average density model} for several values of the inverse temperature $\beta$. Because $\rho_{\beta,2}(z_1)=\rho_{\beta,2}(|z_1|=r_1)$ is rotationally invariant, we only show 1D cuts.  
The two special cases $\beta=2$ and $\beta=0$ are represented by the blue bottom and the red top curve, respectively. For $\beta=2$ the beginning of the formation of a plateau of a constant density on the unit circle can be seen, the circular law, which is known to hold  
in the large-$N$ limit of the Ginibre ensembles (and many other non-Hermitian ensembles, cf. \cite[Chap. 18]{ABD}). 
The 2D Gaussian density in the case of Poisson random variables is also well known.\\

Let us  turn to the nearest-neighbour spacing distribution $p_{\beta,2}(s)$.  Assuming translational invariance, to be justified in Appendix \ref{space}, as in \eqref{pbeta} we condition $z_1=0$ to be at the origin and calculate the probability that the second eigenvalue $z_2$ is at radial distance $s$, 
\begin{eqnarray}
{p}_{\beta,2}(s)&=&\int_\mathbb{C} \mathrm{d}^2z_2\mathcal{P}_{\beta,2}(0,z_2)\delta(|z_2|-s)\nonumber\\
&=&K_\beta \int_0^\infty\mathrm{d}r_2 \int_0^{2\pi} \mathrm{d}\Theta_2 e^{-r_2^2}r_2^{1+\beta}\delta(r_2-s) \nonumber\\
&=&2\pi K_\beta s^{1+\beta} e^{-s^2},
\label{peta'}
\end{eqnarray}
where we used polar coordinates $z_2=r_2e^{i\Theta_2}$. The spacing ${p}_{\beta,2}(s)$ still needs to be normalised and its first moment set to unity. This is achieved by a rescaling \cite{nounfold} as
\begin{equation}
\label{peta}
{p}_{\beta,2}(s)=\frac{2\alpha^{\beta}}{\Gamma[1+\beta/2]}\,s^{1+\beta} e^{-\alpha s^2}, \alpha=\frac{\Gamma[(3+\beta)/2]^2}{\Gamma[1+\beta/2]^2}.
\end{equation}
This is our 
main result for the 2DCG at $N=2$, shown in Fig. \ref{Fig:spacing average density} left for several values of $\beta$.  It correctly reproduces the 2D Poisson distribution (red bottom curve) 
when setting $\beta=0$,
\begin{equation}\label{Poisson2D}
p_{\rm Poisson}^{\rm 2D}(s)= \frac{\pi}{2} s \exp\left[-\frac14\pi s^2\right], 
\end{equation}
which is valid for an infinite number of independent particles , cf. \cite{Haake}. Second, by construction it agrees with the spacing distribution of the GinUE for $N=2$ at $\beta=2$ (blue top curve). 
The spacing of the GinUE is known \cite{GHS}  for any $N$, given in terms of incomplete Gamma functions (or truncated exponentials) as 
\begin{equation}
p_{\beta=2,N}(s)=\prod_{j=1}^{N-1}\frac{\Gamma[1+j,s^2]}{\Gamma[1+j]}\sum_{k=1}^{N-1}\frac{2s^{2k+1}e^{-s^2}}{\Gamma[1+k,s^2]}.
\label{p2Gin}
\end{equation}
In the present form in \eqref{p2Gin} the first moment still needs to be normalised to unity. 
This limiting distribution is universal, cf. \cite{TaoVu,PP}.
Furthermore, it is well known \cite{GHS} that for $N=2$ Eq. \eqref{peta} is {\it not} a good approximation to the limiting spacing distribution of the GinUE \eqref{p2Gin}, 
see however \cite{ABPS} for a surmise for the smallest eigenvalue distribution in non-Hermitian chiral ensembles. 
This is in sharp contrast to 1D, where $N=2$ leads to an excellent approximation of the spacing for $\beta=2$ (and $\beta=1,4$) \cite{Haake}. Thus, one may expect that away from small $\beta$ our spacing \eqref{peta} is not a good approximation 
at large-$N$ either.

\section{Surmise for Spacing Distribution at large $N$ and Applications}\label{Surmise}

Let us show how the spacing \eqref{peta} of a $\beta$-ensemble at $N=2$ can still be used to approximate the spacing distribution of the 2DCG \eqref{2DCoulomb} at large-$N$. The idea here is to apply \eqref{peta} with an {\it effective}, improved value of $\beta_{\rm eff}$, that is determined by a best fit to the spacing in the 2DCG at a given $\beta$. 
In the comparison we use the library of spacings in the 2DCG that was generated numerically in \cite{AKMP} with $N=200$ point charges. We have increased this number to $N=5000$ to improve the statistics of our Coulomb data in the vicinity of $\beta=0$.  The density of the 2DCG is flat for large-$N$, and thus no unfolding is needed
 \cite{Haake,Wettig,AKMP}.  
	\begin{figure}[t!]
		\centering
		\includegraphics[width=0.79\linewidth,angle=0]{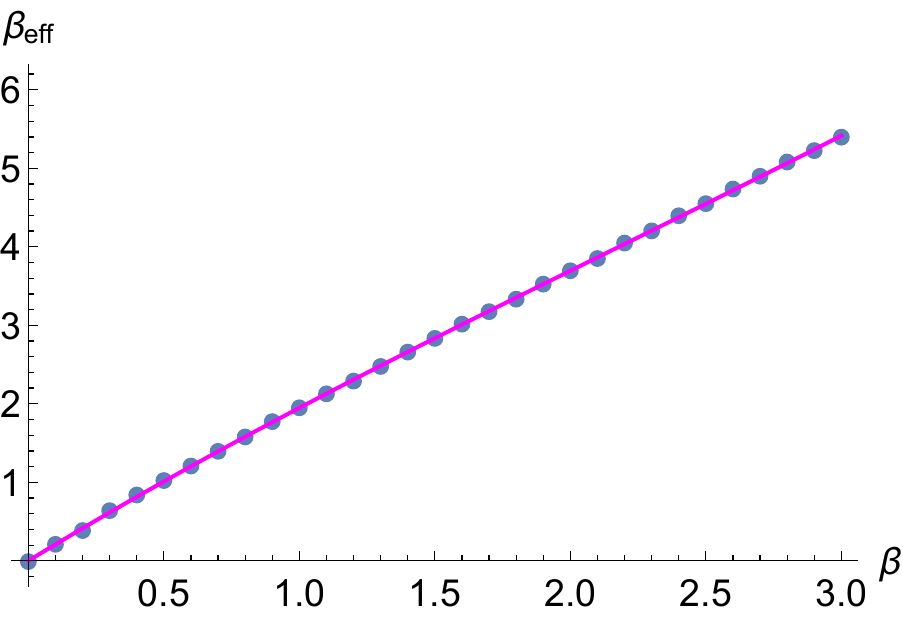}
		\caption{Fit of $\beta_{\rm eff}(\beta)$ with a polynomial of degree 3 from \eqref{cubic} (full line), compared to the best fit $\beta_{\rm eff}$ as a function of $\beta$, in steps of 0.1 (points).}
		\label{Fig:betafit-cubic}
	\end{figure}
The best value for a fitted $\beta_{\rm eff}$ using the functional form \eqref{peta} for the range $\beta\in[0,3]$ is shown in Fig. \ref{Fig:betafit-cubic}, in steps of $\beta$ by 0.1 (points). Although the curve is reasonably well approximated by a straight line, see Appendix \ref{fits} Fig. \ref{Fig:betafit-linear}, we use a polynomial of degree 3, 
\begin{equation}
\beta_{\text{eff}}(\beta)=2.108\beta-0.190\beta^2+0.030\beta^3 ,
\label{cubic}
\end{equation}
as shown in Fig. \ref{Fig:betafit-cubic} above (full line). A discussion comparing with a linear, quadratic and best fit can be found in Appendix \ref{fits}.
Together with
\begin{equation}
\label{petasurmise}
{p}_{\beta}^{\rm surmise}(s)=
\frac{2\alpha_{\rm eff}^{\beta_{\rm eff}}}{\Gamma[1+\beta_{\rm eff}/2]}
\,s^{1+\beta_{\rm eff}} \exp[-\alpha_{\rm eff} s^2],
\end{equation}
where $\alpha_{\rm eff}=\frac{\Gamma[(3+\beta_{\rm eff})/2]^2}{\Gamma[1+\beta_{\rm eff}/2]^2}$, this is our final {\it surmise} for the spacing distribution in the 2DCG or non-Hermitian $\beta$-ensemble. Although it does not capture the expected repulsion $\sim s^{1+\beta}$ at very small values of $s\ll1$, it describes the overall spacing distribution very well, including its maximum and tail.

\begin{figure}[t!]
		\centering
		\includegraphics[width=0.49\linewidth,angle=0]{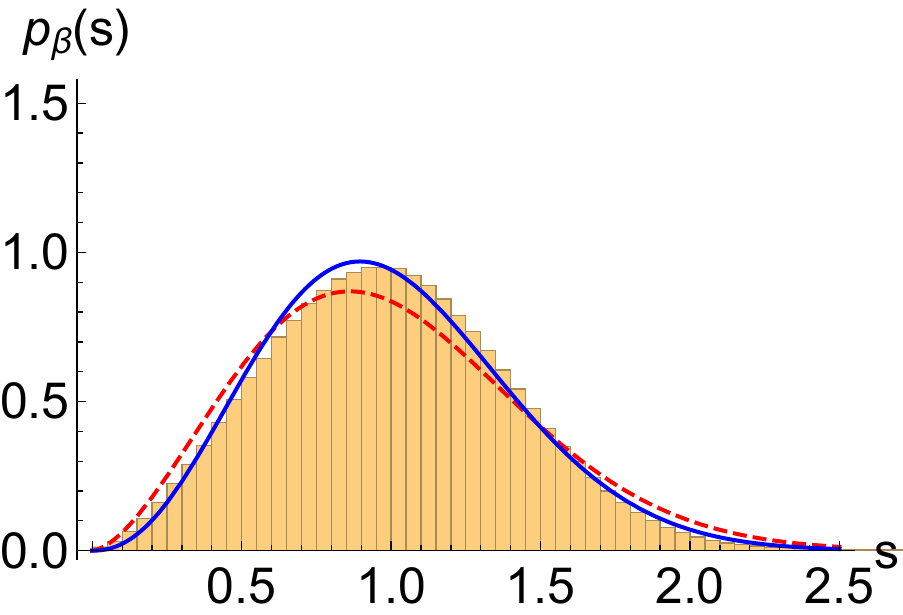}
		\includegraphics[width=0.49\linewidth,angle=0]{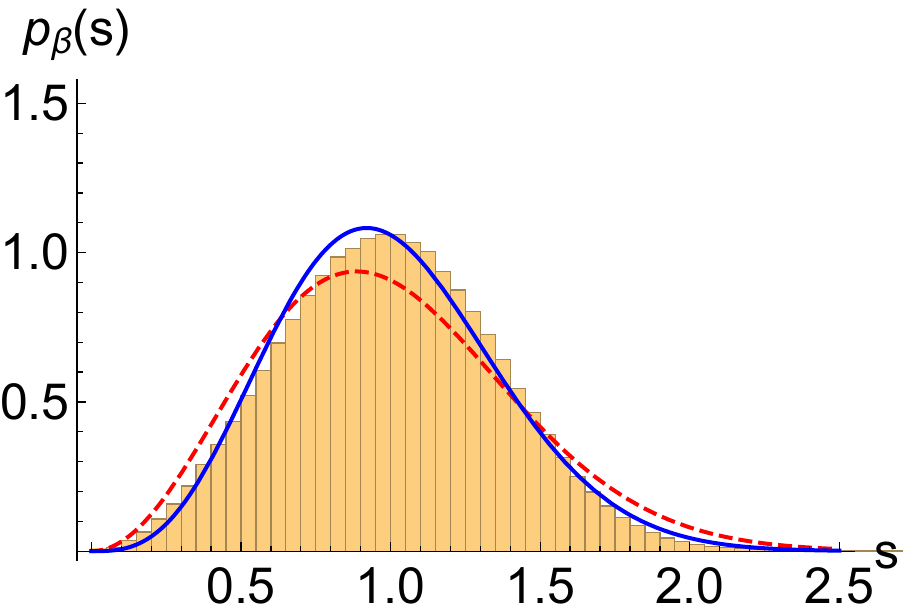}\\
		\hspace{-0.39\linewidth}	\textbf{(a)}\hspace{0.47\linewidth}	\textbf{(b)}\hfill\\
		\includegraphics[width=0.49\linewidth,angle=0]{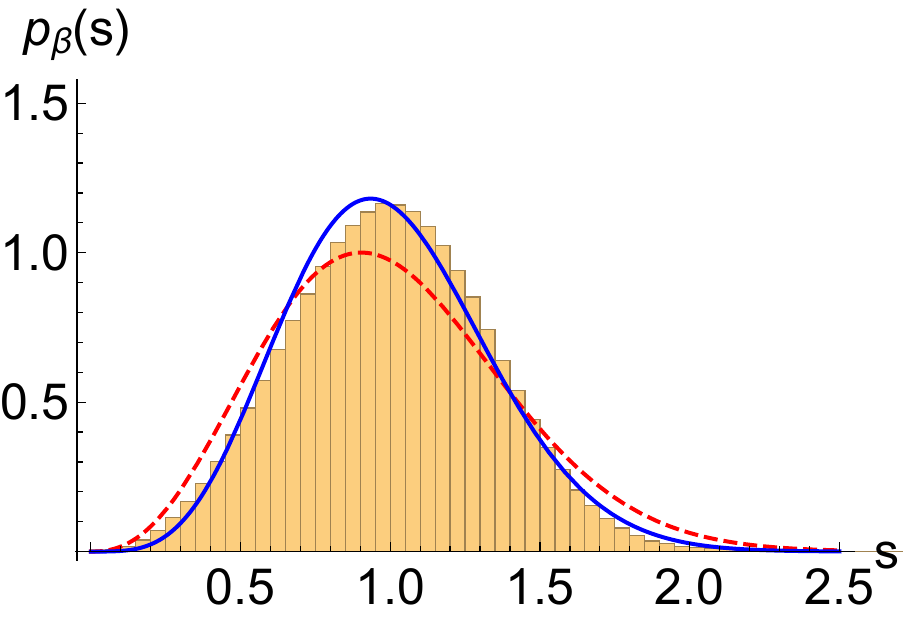}
		\includegraphics[width=0.49\linewidth,angle=0]{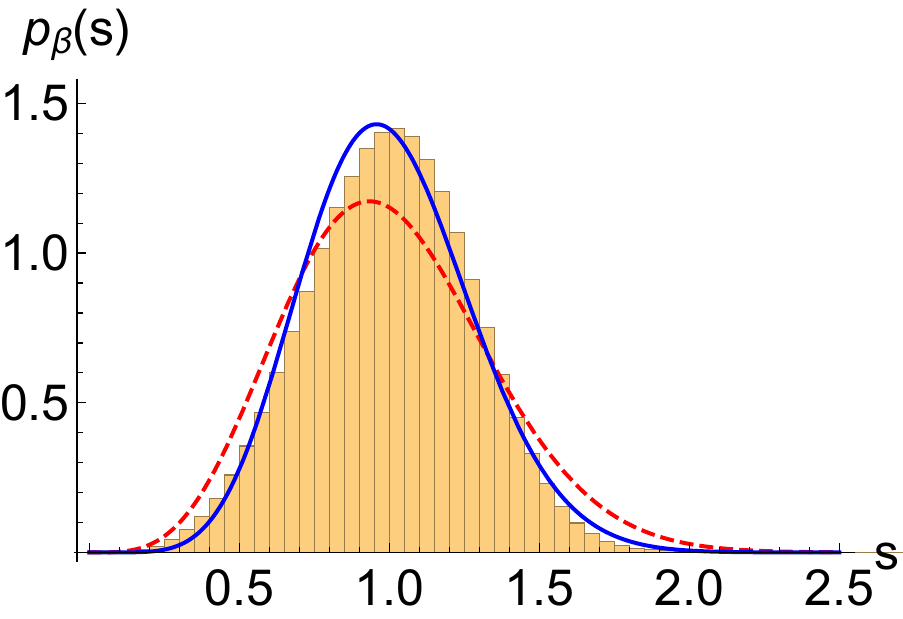}\\
		\hspace{-0.39\linewidth}	\textbf{(c)}\hspace{0.47\linewidth}	\textbf{(d)}\hfill
		\caption{Comparison of 2DCG spacings with $N=5000$ (histograms) at various values of $\beta=0.6,1.0,1.4,2.6$ (plots a)-d)) with our surmise \eqref{petasurmise}  using \eqref{cubic} for $\beta_{\rm eff}$ (blue full line). The $N=2$ result \eqref{peta}  (dashed red line) at the same value $\beta$ {\it without} improvement \eqref{cubic} is not a good approximation.
}
		\label{Fig:pCoulomb}
	\end{figure}

In Fig. \ref{Fig:pCoulomb} we compare our surmise \eqref{petasurmise} to the 2DCG spacing 
for 4 different examples of values of $\beta$ 
in a range of $\beta\in[0,3]$.
Although there is a systematic shift in the maximum, that increases with $\beta$ in the approximate, fitted spacing distribution, the description is surprisingly good for this range of $\beta$. 

The standard deviation $\sigma$, that is used for the fits shown in Fig. \ref{Fig:betafit-cubic}, is presented in Table \ref{Tab.KS} at the fit points. It is defined in the usual way for a chosen $\beta$ and fitted value $\beta_{\rm eff}$:
\begin{equation}
\sigma=\left[ \frac{1}{n}\sum_{j=1}^n(p_{\beta}(s_j)-p_{\beta}^{\rm surmise}(s_j))^2
\right]^{\frac12},
\label{sigma-def}
\end{equation}
where $n$ is the number of bins in the data and $p_\beta(s_j)$ the number of counts in the $j$-th bin at its mid point $s_j$. In all data the sum was cut off at $s=3$, where we have an exponential suppression.
To estimate the systematic error for our size of ensemble, we have compared GinUE eigenvalues of matrix size $N=5000$ and the 2DCG at the same value $\beta=2$ and $N$, leading to $\sigma=0.44\cdot 10^{-2}$, 
and Kolmogorov-Smirnov distance  $d=0.20\cdot 10^{-2}$. 
Here, $d$ is defined as 
\begin{equation}
\label{KS-def}
d=\max_{x\geq0}|F(x)-G(x)|\in[0,1],
\end{equation}
between the cumulative distributions $F$ and $G$ of distributions $f$ and $g$, that is of the spacing distribution of the 2DCG and surmise here. 
The Kolmogorov-Smirnov distance $d$ is independent of the binning into histograms. It remains almost constant from $\beta=1.9-3.0$, see Table \ref{Tab.KS}, where we show its values along with $\sigma$. 

\begin{table}[t!]\centering
	\caption[]{List of standard deviations $\sigma$ and Kolmogorov-Smirnov distances $d$ (both in units $ 10^{-2}$) 
	between fitted surmise \eqref{petasurmise} using \eqref{cubic} and 2DCG,  
	with $\beta\in[0,3]$ in steps $0.1$.}\vspace{3mm}
		\begin{tabular}{|c|c|c|c|c|c|c|c|c|c|c|c|c|c|c|c|}
			\hline
			$\beta$  & 0.1  & 0.2 & 0.3 & 0.4 & 0.5 & 0.6 & 0.7 & 0.8& 0.9  &1.0 &1.1&1.2&1.3&1.4&1.5\\
			\hline\hline
			$\sigma$ & 0.9  & 1.4 & 1.9 & 2.3 & 2.6 & 2.8 & 3.1 & 3.4 &3.6  &3.7 & 3.8 & 4.0  & 4.1 & 4.2 & 4.3\\
			\hline
			$d$ & 0.7  & 0.6 & 0.9 & 1.1 & 1.3 & 1.5 & 1.6 & 1.7 &1.8  &1.9&2.0 & 2.0  & 2.1 & 2.1 & 2.1 \\
			\hline
			\hline
			$\beta$ & 1.6&1.7&1.8&1.9&2.0& 2.1 & 2.2  & 2.3 & 2.4 & 2.5 & 2.6 & 2.7 & 2.8 & 2.9& 3.0  \\
			\hline\hline
			$\sigma$ & 4.3 & 4.5 &4.5 & 4.6& 4.7 & 4.9 & 4.9  & 4.9 & 4.9 & 5.2 & 5.3& 5.2 &5.2 & 5.4& 5.4   \\\hline
			$d$ &  2.2 & 2.2 &2.2 & 2.3& 2.3 & 2.4 & 2.3  & 2.3 & 2.4 & 2.4 & 2.4& 2.4 &2.4 & 2.4& 2.4    \\
			\hline
		\end{tabular}
	\label{Tab.KS}
\end{table}

In contrast to the Wigner surmise in 1D \eqref{pWigner}, which becomes a better approximation for increasing values of $\beta=1$ to $\beta=4$, but fails when getting closer to 1D Poisson \eqref{Poisson1D}
at lower $\beta\to0$, we observe the opposite behaviour here. The closer we get to 2D Poisson \eqref{Poisson2D}, the better is the approximation, see Tab. \ref{Tab.KS} and Fig. \ref{Fig:pCoulomb}. Thus when quantifying the transition between integrable (Poissonian) behaviour, fully chaotic random matrix statistics in the respective symmetry class, 
and potential intermediate transitions, 
our surmise may serve as a good parametrisation. 

	\begin{figure}[h!]
		\centering
				\includegraphics[width=0.49\linewidth,angle=0]{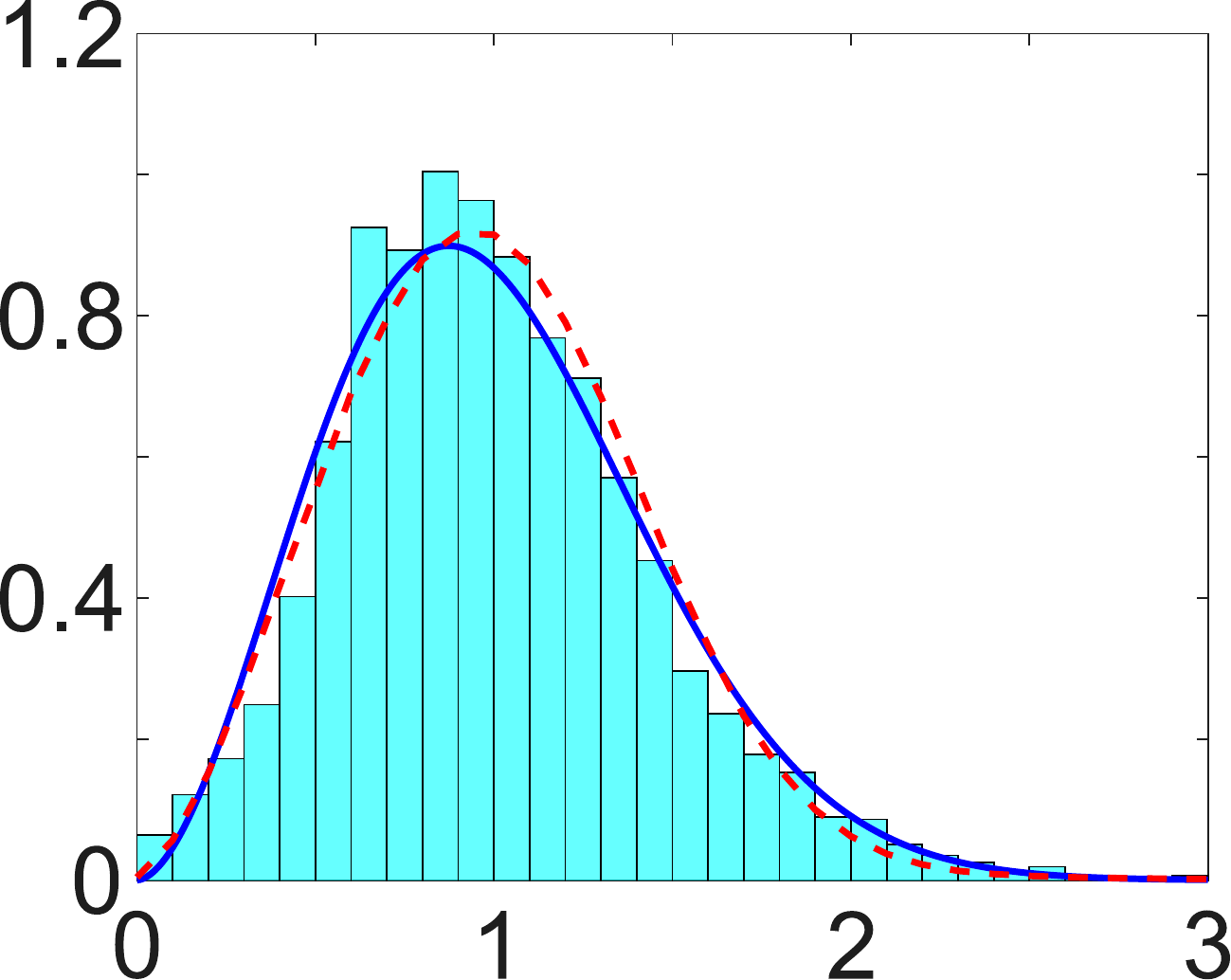}
		\includegraphics[width=0.49\linewidth,angle=0]{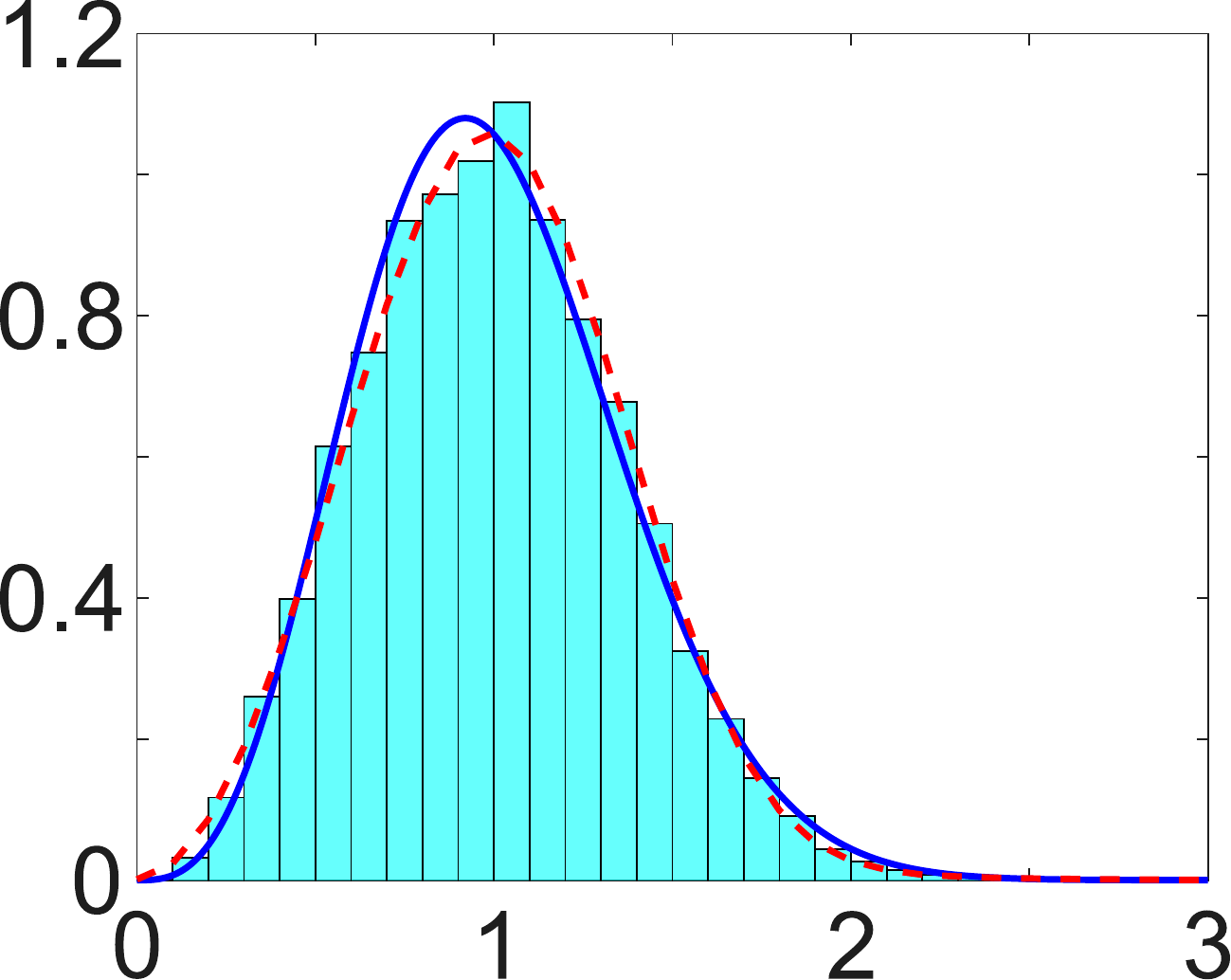}
		\caption{Left: Fit of
		the spacing distribution between occupied Buzzards nests (histograms)
observed annually from 2000-2019 \cite{ABC} with surmise \eqref{petasurmise} at $\beta=0.38$ (red dashed line) and 2DCG at $\beta=0.5$ (blue full line). Right: 
Fit of the spacing between the complex eigenvalues of the Liouville operator of an XXX quantum spin chain \cite{AKMP}  (histograms) by  $\beta=1(0.99)$ from the 2DCG in blue (surmise in dashed red). 
}
		\label{Fig:Liouville}
	\end{figure}

In order to illustrate the usefulness of our surmise we test it on real data from two published works \cite{ABC,AKMP} in such an intermediate range for $\beta$, and compare with a fit to the numerically generated 2DCG. 
Our first example comes from theoretical Biology, where the 2DCG is used as a simple statistical mechanics model to describe the behaviour of birds \cite{ABC}.
In Fig. \ref{Fig:Liouville} left, the spacings between occupied nests of the common buzzard in an area of the Teutoburger forrest are shown, where every year serves as a separate ensemble. The non-Hermitian $\beta$-ensemble is used as a simple, one-parameter model to quantify the repulsion between these highly territorial birds of prey, see \cite{ABC} for more discussion. Taking all $3135$ 
data points from the entire period of observation over 20 years, we obtain $\beta=0.38(0.5)$ 
{\bf check}
from our surmise (2DCG). Furthermore, in \cite{ABC} a time moving averages of 5 years was used to detect a change in repulsion as measured by $\beta$ over time, with about $500$ data points per time window. On these smaller data sets we have observed similar or larger deviations  of about 20\% in the obtained $\beta$-value, comparing the 2DCG and our surmise. Notice however, that the fitted $\beta$ does not have a biological meaning. It serves as a simple way to quantify time dependence and correlation length (as measured by the next-to-nearest-neighbour spacing in \cite{ABC}) in the observed repulsive behaviour. Thus the change in $\beta$ is meaningful, and not its actual value at a specific instance of time.

The second example comes from Physics, where we investigated a boundary driven  XXZ quantum spin-chain. Different choices of parameters allow to study the transition between integrable and chaotic behaviour on the spectrum which is complex, cf. \cite{AKMP}.
Fig. \ref{Fig:Liouville} right shows a comparison with 77520 spacings of the complex eigenvalues of the Liouville operator of an isotropic Heisenberg XXX spin chain where only the zero-mode is integrable \cite{Tomaz}, cf. \cite{AKMP} for details of the parameters chosen and a discussion.  Due to the high quality of data, a very precise fit of $\beta$ can be made, that shows very little deviation between our surmise and the 2DCG. Increasing the parameters responsible for the dissipation, in \cite{AKMP} an agreement with the GinUE at $\beta=2$ \eqref{p2Gin} was found, which is consistent with fully chaotic behaviour.

\section{
Spacing Distributions in Classes AI$^\dagger$ and AII$^\dagger$ Approximated by 2DCG
}
\label{SymmClass}
Here, we study the spacing distribution of the two non-Hermitian symmetry classes AI$^\dag$ and AII$^\dag$ of random matrices. As it was pointed out in \cite{KSUS} using  numerical simulations, these are the only two among the 38 classes that show a different behaviour from the spacing distribution of the Ginibre ensemble in the bulk of the spectrum. While the latter is know, see \eqref{p2Gin}, no analytic or approximate form for the spacing in these two classes is known.
We use the nomenclature of \cite{HKKU} which can be clarified as follows: Let $J\in \mathbb{C}^{N^2}$ be a complex non-Hermitian $N\times N$ matrix. The complex Ginibre ensemble is denoted by 
\begin{equation}
	\label{defA}
	\mbox{\rm class A}: \quad J.
\end{equation}
The ensemble of real, non-symmetric matrices without further symmetry constraints is called 
\begin{equation}
	\label{defAI}
	\mbox{\rm class AI}: \quad J=J^*,
\end{equation}
with $J^*$ the complex conjugate of $J$.
Furthermore, the ensemble of matrices with quaternion entries defines the symmetry class AII. If we choose the complex representation of such $N\times  N$ matrices of size $2N$, we obtain 
\begin{equation}
	\label{defAII}
	\mbox{\rm class AII}:J=\Sigma J^* \Sigma, 
\end{equation}
with the skew metric
\begin{equation}
	\Sigma=\left( \begin{matrix}
		0 & -i1_{N\times N}\\
		i1_{N\times N} & 0\\
	\end{matrix} \right).
\end{equation} The distribution of these three Ginibre ensembles is Gaussian, cf. \eqref{betaRM} at $\beta=2$ ($\eta=0$):
\begin{equation}
	\mathcal{P}_{2,N}(J)\propto \exp[-\Tr(JJ^\dag)].
	\label{Gauss}
\end{equation} 
In the study of non-Hermitian matrices we obtain in general a different symmetry class, if we consider the transpose on the right hand side instead of the complex-conjugate.  Therefore, class AI$^\dag$ is defined following class AI, where on the right hand side of the defining equation \eqref{defAI} $J^*$ is replaced by the transposed matrix $J^T$. It describes the class of \textit{complex symmetric matrices}:
\begin{equation}
	\label{defAI dag}
	\mbox{\rm class AI}^{\dag}: \quad J=J^T.
\end{equation}
Likewise, we obtain the class AII$^\dag$ of \textit{complex self-dual quaternion} matrices by the definition
\begin{equation}
	\label{defAII dag}
	\mbox{\rm class AII}^{\dag}:J=\Sigma J^T \Sigma. 
\end{equation}
Also in these ensembles the distribution of matrix elements is Gaussian, given by \eqref{Gauss}. 
It was pointed out in \cite{Hastings} that the ensemble AII$^\dag$ can be realised by antisymmetric complex matrices $A=-A^T$ via $J=i\Sigma A$, and in that sense these two ensembles are related. As far as we are aware of,  no results are known to date about the joint densities of complex eigenvalues of classes AI$^\dag$ and AII$^\dag$, including for $N=2$.  Using a saddle point approximation, it was shown in \cite{Hastings} that for large eigenvalue separation class  AII$^\dag$ is approximately described by a 2DCG at $\beta=4$.

\subsection{Comparing analytic results at $N=2$}
\begin{figure}[b!]
	\centering
	\includegraphics[width=0.49\linewidth,angle=0]{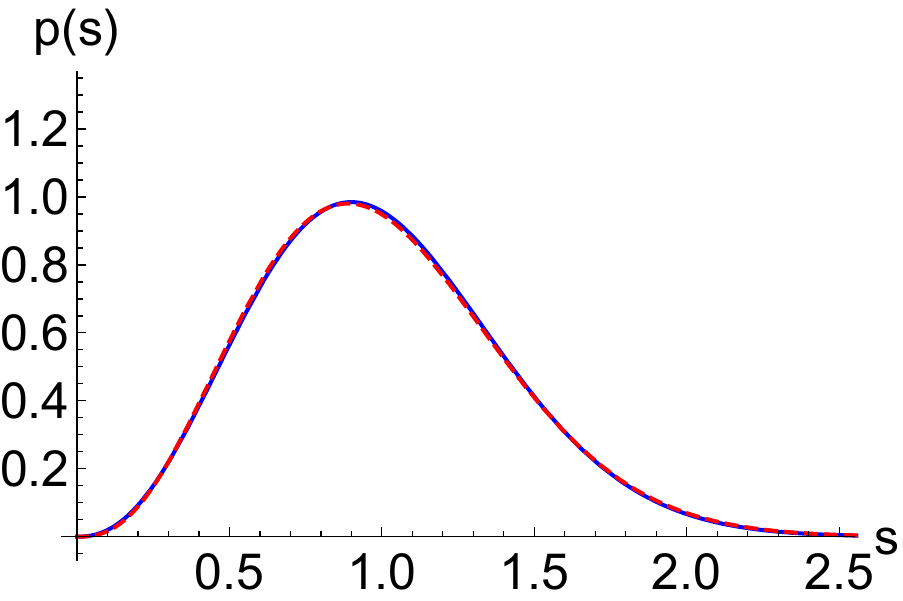}
	\includegraphics[width=0.49\linewidth,angle=0]{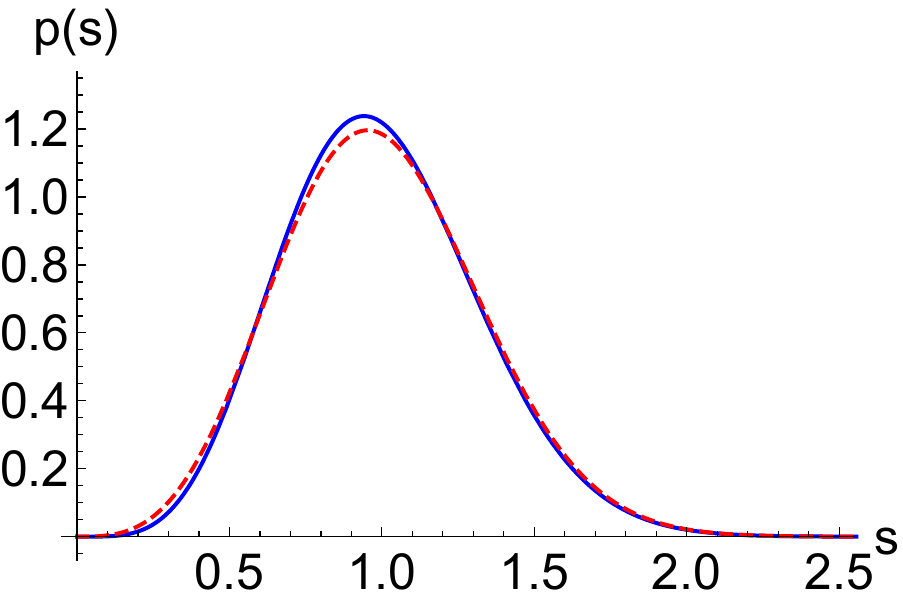}
	\caption{Comparison of the spacing distribution at  $N=2$ 
		for class AI$^\dag$ \eqref{pAI} (red dashed line)	with the $\beta$-ensemble \eqref{peta} (blue full line) at $\beta=1.3$ with standard deviation $\sigma=0.57\cdot 10^{-2}$ and $d=0.34\cdot10^{-2}$
		(left),  and class AII$^\dag$  \eqref{pAII} (red dashed line) with \eqref{peta}  (blue full line) at $\beta=2.8$ 
		with $\sigma=1.16\cdot 10^{-2}$ and $d=0.58\cdot10^{-2}$ (right).		The first moment of all spacings is normalised to unity.
	}
	\label{Fig:compareN2}
\end{figure}

For $N=2$, starting from the matrix representation the spacing distribution could be  computed analytically \cite{HKKU}, see also \cite{JPP} for AI$^\dag$.
\begin{eqnarray}
	p_{{\rm AI}^\dag,N=2}(s)&=& 2C_{\rm AI^\dag}^4 s^3 K_0(C_{\rm AI^\dag}^2s^2),
	\label{pAI}
	\\
	p_{{\rm AII}^\dag,N=2}(s)&=& \frac23 C_{\rm AII^\dag}^4 s^3\left(1+C_{\rm AII^\dag}^2s^2\right)e^{-C_{\rm AII^\dag}^2s^2},\quad
	\label{pAII}
\end{eqnarray}
where $C_{\rm AI^\dag}=\Gamma[1/4]^2/2^{7/2}$ and $C_{\rm AII^\dag}=7\sqrt{\pi}/8$. Both spacing distribution are normalised to unity, including their first moment, and are shown in Fig. \ref{Fig:compareN2}.
Here, $K_0(x)$ denotes the modified Bessel-function of the second kind. 

For completeness we briefly recall the derivation of eqs. \eqref{pAI} and \eqref{pAII}, without determining the joint density of eigenvalues.
Notably for class AI$^\dag$ it was used in \cite{JPP} that the two complex eigenvalues can be computed explicitly in terms of the matrix elements $J_{kl}$, yielding the expression $s=|z_2-z_1|=\sqrt{|(J_{11}-J_{22})^2-4J_{12}J_{21}|}$. The integration over the matrix elements can then be carried out.
The main idea of the proof in \cite{HKKU} is to argue that the spacing $s$ is proportional to the random variable $\sqrt{|\chi_f|}$, with $\chi_f=y_1^2+\ldots+y_f^2$, for $y_1,\ldots,y_f$ independently and identically Gaussian distributed complex variables. Here, we have $f=2$ for class AI$^\dag$, $f=3$ for class A, and $f=5$ for class AII$^\dag$, depending on the number of basis elements in the corresponding ensemble. First, one can calculate the probability distribution function of $|\chi_f|^2$
\begin{equation}
	p(|\chi_f|^2=\rho)\propto\int_{\mathbb{C}^{f}}\mathrm{d}^{2f} \vec{y} e^{-||\vec{y}||^2}\delta(\rho-|y_1^2+\ldots+y_f^2|^2).
\end{equation}
This integration can be performed after some change of variables, using multi-dimensional spherical coordinates. After an appropriate change of variables 
$s=\rho^{1/4}$, the final distribution then reads
\begin{align}
	p\left(s=\sqrt{|\chi_f|}\right) \propto s^{f+1} K_{\frac{f}{2}-1}(s^2),
\end{align}
where $K_\nu(x)$ is again the modified Bessel-function of the second kind. Using its properties for half-integer values of $\nu$, cf. \cite{Gradshteyn}, leads to \eqref{pAI}, \eqref{pAII} and \eqref{p2Gin} at $N=2$. The corresponding normalisation was also computed in \cite{HKKU,JPP}.

Next, let us look at the behaviour of the spacing \eqref{pAI} and \eqref{pAII} for asymptotically small $s\ll 1$ and large $s\gg 1$. For class AI$^\dag$ we find
\begin{eqnarray}
	p_{{\rm AI}^\dag,N=2}(s)\sim \begin{cases}
		s^3\left[\ln (s)(1 +\mathcal{O}(s^4))
				\right],&  s\ll 1\\
		s^2e^{-C_{\rm AI^\dag}^2s^2} \left[1+\mathcal{O}\left(\frac{1}{s^2}\right) \right],& s\gg 1,\nonumber
	\end{cases}
\end{eqnarray}
where we used for the asymptotic of $K_0(z)$ for $z\ll 1$ \cite[8.447]{Gradshteyn}
\begin{eqnarray}
	K_0(z)=&-\ln(\frac{z}{2})I_0(z)+\sum_{k=0}^\infty \frac{z^{2k}}{2^{2k}k!}\psi(k+1)\nonumber\\
	=&-\ln(\frac{z}{2})(1+\mathcal{O}(z^2))-\gamma + \mathcal{O}(z^2), 
\end{eqnarray}
with $\psi(1)=-\gamma$ the Euler-Mascheroni constant \cite[8.3366(1)]{Gradshteyn}. The large argument asymptotic of $K_0(z)$ for $z\gg 1$ follows from  \cite[8.451(6)]{Gradshteyn}
\begin{equation}
	K_0(z)=\sqrt{\frac{\pi}{2z}}e^{-z}\left[1+\mathcal{O}\left(\frac{1}{z}\right) \right].
\end{equation}
For class AII$^\dag$ we can directly read off the asymptotic behaviour from \eqref{pAII}. Compared to the spacing in the GinUE at $N=2$ from \eqref{p2Gin}, $p_{2,2}(s)=2s^3 e^{-s^2}$,
class AII$^\dag$  in \eqref{pAII} shares the same cubic repulsion $\sim s^3$ at $s\ll1$. This  feature was shown in \cite{Haake} to hold universally for a large class of ensembles at large-$N$, using perturbation theory. In contrast, due to the logarithmic singularity at the origin in \eqref{pAI},  class AI$^\dag$ behaves rather as $\sim s^3\ln(s)$. All three spacing distributions have a Gaussian tail for $s\gg1$.
Let us emphasise that, as it is already known for the GinUE \cite{GHS}, neither \eqref{pAI} nor \eqref{pAII} at $N=2$ provide good approximations for the respective limiting spacing distributions at large $N$, see Fig. \ref{Fig:fitAI-II} below.

As a first step,  and in the same spirit as in the previous Section \ref{Surmise}, let us see if the analytical results \eqref{pAI} and \eqref{pAII} at $N=2$  can be  approximated by our spacing distribution $p_{\beta,2}(s)$ of a 2DCG also at $N=2$ \eqref{peta}, by fitting $\beta$ as shown in Fig. \ref{Fig:compareN2}. Despite the different logarithmic behaviour at very small spacing for class AI$^\dag$, the approximation with a non-integer value $\beta=1.3$ is excellent for this class. The precise values of the  fitted $\beta$ 
given in Fig. \ref{Fig:compareN2} 
are not so important here, as we are not yet in the large-$N$ limit, which is supposed to be universal \cite{HKKU}. As we have seen in the main text, such an approximation by a 2DCG gas pertains also there, as explained now.

\subsection{Spacing at large $N$: Approximation by non-integer $\beta$-Ensembles}
Let us approximate the spacing distribution of  symmetry classes AI$^\dag$ and AII$^\dag$ of non-Hermitian random matrices at large-$N$ by a 2DCG with fitted $\beta$. Their spacing distributions are unknown beyond $N=2$ \cite{HKKU,JPP}, and their large-$N$ limit is found to be universal from numerical results with matrices with Bernoulli distribution \cite{HKKU}.

We aim at a precise determination of the fitted values of $\beta$ in these two non-Hermitian classes AI$^\dag$ and AII$^\dag$ for large-$N$, relying on the 2DCG directly and not our surmise, that looses precision for increasing values of $\beta$. Notice, that for Hermitian random matrices the Dyson index $\beta$ as it appears in \eqref{pWigner} is directly given by the number of independent real degrees of freedom per matrix element. 
	\begin{figure}[b!]
	\centering
	\includegraphics[width=0.39\linewidth,angle=0]{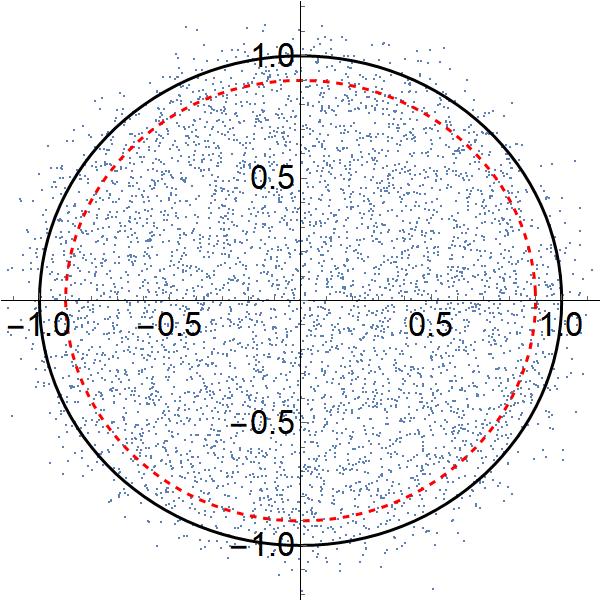}
	\includegraphics[width=0.39\linewidth,angle=0]{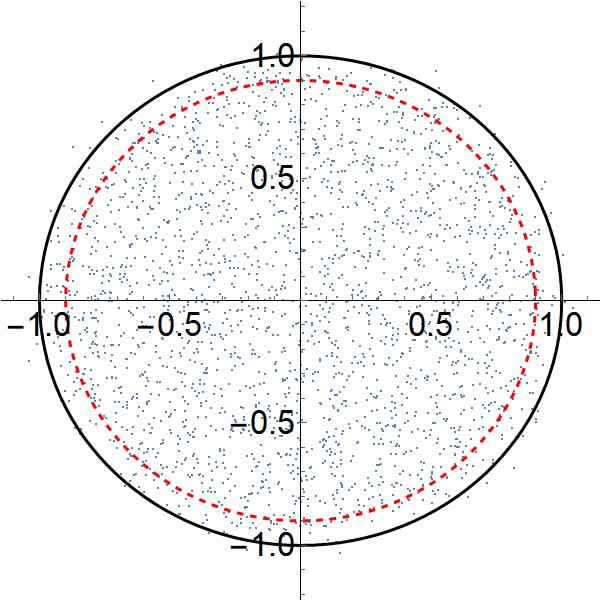}
	\caption{Scatterplot of complex eigenvalues of 50 complex symmetric matrices (AI$^\dag$) of size $N=100$ (left), and of 50 complex quaternion self-dual matrices (AII$^\dag$) of size $2N=100$ (right), where every eigenvalue (point) is doubly degenerate. To ensure that we probe the bulk statistics, 
	only the spacings among eigenvalues inside the red rings are used. It radius is chosen approximately at equal distance from the limiting support (black curve) as the eigenvalues lying outside of it.  Both averaged densities can be seen to be constant inside the considered region.}
	\label{Fig:scatter}
\end{figure}

Because the density of the ensembles AI$^\dag$ and AII$^\dag$ is constant in both cases for large-$N$, see Fig. \ref{Fig:scatter}, no unfolding is needed here, see \cite{Wettig,AKMP,Haake} for a discussion of this issue. Second, as it is well known from the complex Ginibre ensemble \cite{GHS}, the correlations among eigenvalues change close to the edge, compared to the bulk of the spectrum. Thus, in order to probe bulk statistics, we only consider the spacing among eigenvalues within the red ring away from the edge shown in Fig. \ref{Fig:scatter} (the nearest neighbour may lie outside the red ring though). 

\begin{figure}[t!]
	\centering
	\includegraphics[width=0.49\linewidth,angle=0]{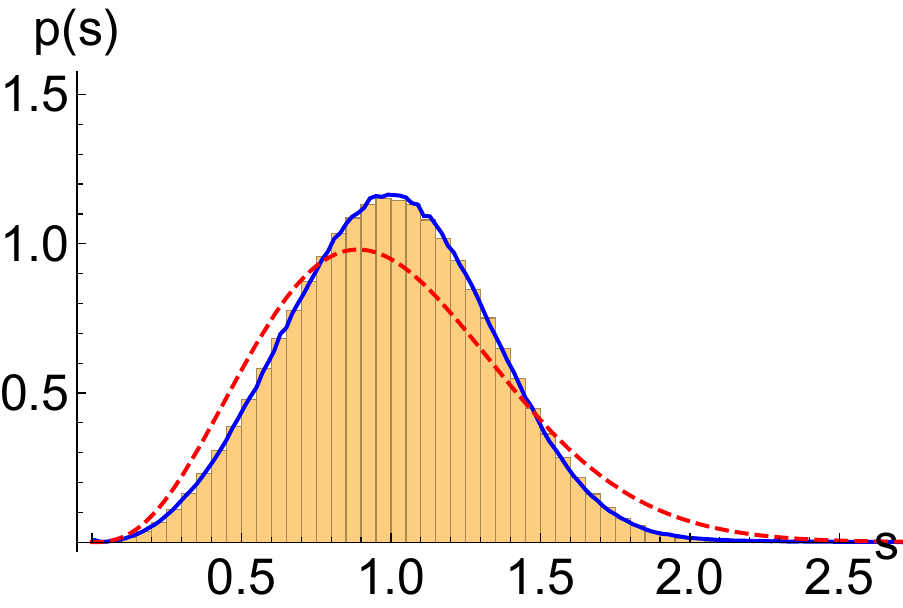}
	\includegraphics[width=0.49\linewidth,angle=0]{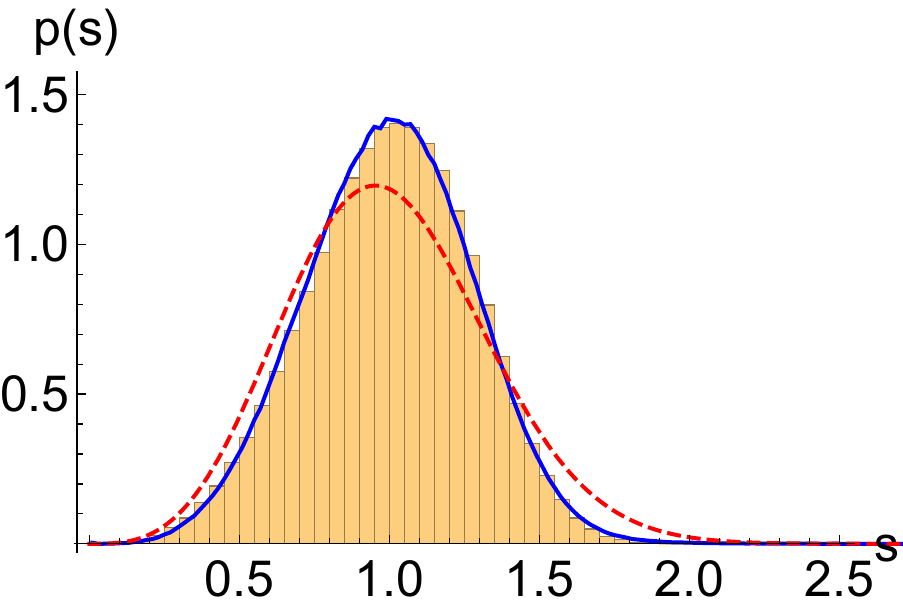}
	\caption{Comparison  of the spacing distribution $p(s)$ of complex eigenvalues for 500 matrices of size $N=5000$
		for class AI$^\dag$ with \eqref{pAI} at $N=2$ (red dashed line), and with the 2DCG  (blue full line) at $\beta=1.4$, with $\sigma=0.86\cdot10^{-2}$ and $d=0.26\cdot10^{-2}$	(left). For class AII$^\dag$ we compare with  \eqref{pAII} at $N=2$ (red dashed line), and with the 2DCG  (blue full line) at $\beta=2.6$, with 
$\sigma=2.29\cdot10^{-2}$ and $d=0.91\cdot10^{-2}$			(right). The first moment of all spacings is normalised to unity. 
	}
	\label{Fig:fitAI-II}
\end{figure}

For the generation of the spacing distribution of the  classes AI$^\dag$ and AII$^\dag$ we considered  500 random matrices of size $N=5000$, respectively $2N=5000$. Figure  \ref{Fig:fitAI-II} shows that the spacing distribution of both classes finds an excellent approximation by a 2DCG at fitted {\it non-integer value} of $\beta=1.4$ for class AI$^{\dag}$, and $\beta=2.6$ for class AII$^\dag$. We have used step size $0.1$ in fitting $\beta$. The best fit to the 2DCG is shown by the blue curve, which describes both tails and the global maximum well. We use the approach presented in \cite{AKMP} and extended the 2DCG to $N=5000$  points as well, to improve our precision. 
For comparison, the red dashed curves show the corresponding spacing distributions 
\eqref{pAI} and \eqref{pAII} at $N=2$,  which clearly do not approximate the spacing at large-$N$ well.

\begin{figure}[b!]
	\centering
	\includegraphics[width=0.49\linewidth,angle=0]{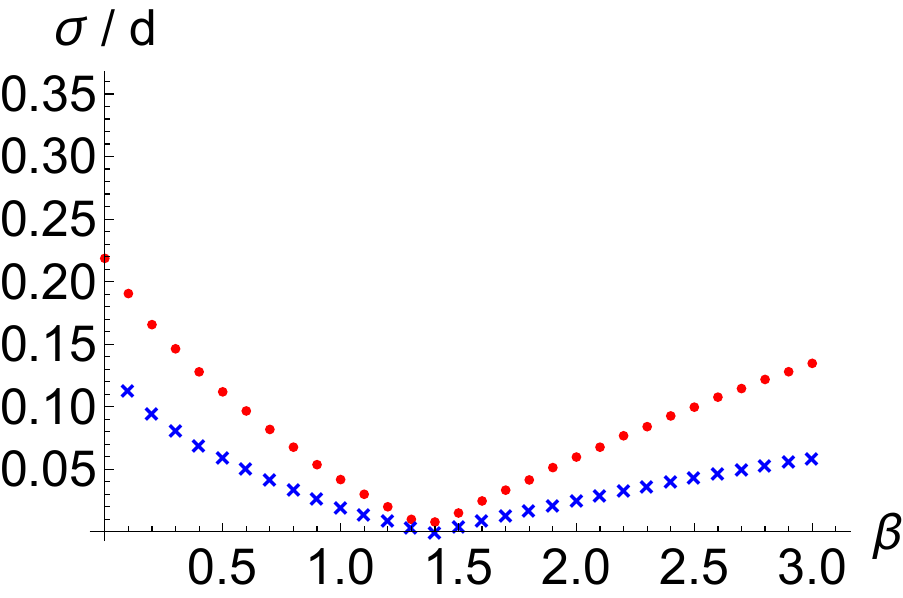}
	\includegraphics[width=0.49\linewidth,angle=0]{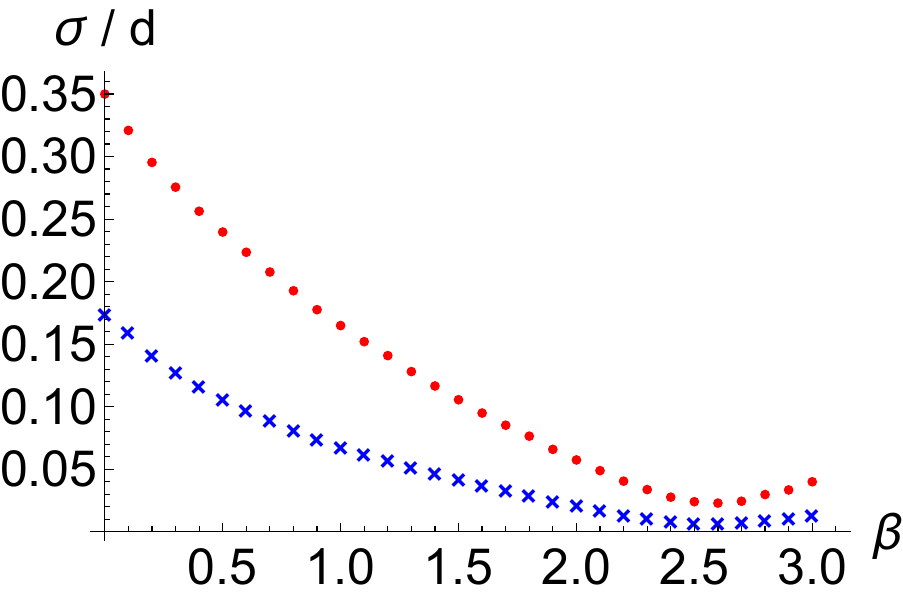}
	\caption{Standard deviation $\sigma$ (red dots) and Kolmogorov-Smirnov distances $d$ (blue crosses) between the 2DCG data at different values of $\beta\in[0,3]$, and the bulk spacings for random matrices of class AI$^\dag$ (left) and AII$^\dag$ (right) from Fig. \ref{Fig:fitAI-II}.}
	\label{Fig:betafit}
\end{figure}

In Fig. \ref{Fig:betafit} we quantify the quality of our fit by plotting the standard deviation $\sigma$ \eqref{sigma-def} and Kolmogorov-Smirnov distance $d$ \eqref{KS-def}, 
respectively, obtained at different values of $\beta$ of the 2DCG, in step size $0.1$. The best fit is indicated by the local minimum and agrees for both measures of distance. From the plots it seems that the closest half integer values of $\beta=1.5$ for AI$^\dag$ (left) and of $\beta=2.5$ for AII$^\dag$ (right) are disfavoured. We should keep in mind however, that the fit by the 2DCG is an effective description of the otherwise unknown local statistics in these two symmetry classes.

Let us emphasise that, apart from the spacing distribution at $N=2$, so far no further local information about complex eigenvalue correlations in class AI$^\dag$ and AII$^\dag$ is available. Currently we have no heuristic  explanation for the fitted $\beta$-values. With the 2DCG description it is easy to compare complex eigenvalue data with a given symmetry class, as it was done e.g. for spin chains \cite{HKKU} and the kicked rotor \cite{JPP}, without the need to diagonalise ensembles of large random matrices. 
As we mentioned already, in \cite {Hastings} an approximate joint distribution of complex eigenvalues was  obtained for the symmetry class AII$^\dag$, that  becomes proportional to $|\Delta_N(Z)|^4$ for large separation, that is a  2DCG at $\beta=4$. We find, however, that the local repulsion is very different in this class. If we were to speculate that the spacing in  AI$^\dag$ remains $\sim s^3\ln[s]$, and in AII$^\dag$ remains $\sim s^3$ for small separation $s\ll1$ at large-$N$, clearly their spacing does not follow that of the Ginibre ensemble at $\beta=2$, that shares  $\sim s^3$ for $s\ll1$ at large-$N$. This would mean that the local cubic repulsion at $s\ll1$, claimed in \cite{Haake} perturbatively for a large class of ensembles, does not tell us much about the overall form of the spacing distribution.

\section{Conclusions}

In this 
paper we have analysed the nearest neighbour spacing distribution in the complex plane resulting from a 2D Coulomb gas at inverse temperature $\beta$ in a confining potential. We have proposed an invariant matrix representation based on $N$-dimensional complex normal random matrices. The case $N=2$ with a Gaussian potential has led us to formulate a surmise, after introducing an effective $\beta_{\rm eff}$ in a cubic fashion.
It works very well for small values of $\beta$ close to 2D Poisson. This is in contrast to the Wigner surmise in 1D which applies to $\beta$ of order unity or larger. 
This approximation allowed us to provide a simple measure to study the transition in $\beta$ between uncorrelated and random matrix statistics in different symmetry classes. The transition has been illustrated by a comparison to spacing distributions from data in a quantum spin chain and ecology. Furthermore, we have established an approximate 2D Coulomb gas description of the spacing distribution in the two symmetry classes of non-Hermitian random matrices AI$^\dag$ and AII$^\dag$ at non-integer values of $\beta=1.4$ and $\beta=2.6$, respectively, compared to the complex Ginibre ensemble at $\beta=2$. Although the situation is reminiscent to the 3 classes of Hermitian random matrices with $\beta=1,2,4$, it remains an open problem to explain these non-integer values, and to provide further local spectral information in these two symmetry classes. 
An alternative matrix representation for the non-Hermitian $\beta$-ensemble, perhaps  analogous to the tridiagonal parametrisation of Dumitriu-Edelman in the Hermitian $\beta$-ensemble, would be highly desirable to find as well.

\paragraph*{Acknowledgements.} 
This work was funded by the Deutsche Forschungsgemeinschaft (DFG) grant SFB 1283/2 2021 – 317210226 "Taming uncertainty and profiting from randomness and low regularity in analysis, stochastics and their applications" (GA and PP). We thank Mario Kieburg for useful discussions.

\begin{appendix}
\section{Non-Hermitian $\beta$-Ensemble at $N=2$}\label{Sec.BN2}

In this appendix we provide some details about the non-Hermitian ensemble of Gaussian $2\times2$ random normal matrices, defined in Eq. \eqref{betaRMN2} in the main text. In particular we compute the normalisation constants of the distribution of matrix elements and distribution of eigenvalues in Subsection \ref{jpdf}. Subsection \ref{space} is devoted to the derivation of the spacing distribution without assuming translational invariance.

\subsection{Normalisation constants}\label{jpdf}

We compute the normalisation constants $C_\beta$ and $K_\beta$ of the distribution of matrix elements and complex eigenvalues, respectively, starting with the former. From the parametrisation \eqref{parametrization} of the complex normal matrix $J$, we obtain the following distribution by inserting it into \eqref{betaRMN2}:
\begin{eqnarray}
\mathcal{P}_{\beta,2}(a,b,k_1,k_2)
&=&\frac{e^{-2(|a|^2+|b|^2+k_2(a_1\cos k_1+a_2\sin k_1))-k_2^2}}{C^{-1}_\eta(4|b|^2+k_2^2)^\eta},
\nonumber\\ 
	C^{-1}_\eta&=&\frac{\pi^3}{2^{\eta+1/2}}\Gamma\left[\frac32-\eta\right], 
\label{betaRMN2'}
\end{eqnarray}
where we keep the parameter $\eta=1-\beta/2$ in the remainder of this appendix. We need to show that the integral of \eqref{betaRMN2'} is normalised:
\begin{equation}
1=
\int_{0}^{+\infty}\mathrm{d}k_2 \int_{0}^{2\pi}\mathrm{d}k_1
\int_{\mathbb{C}}{\rm d}^2a \int_{\mathbb{C}}{\rm d}^2b \ 
\mathcal{P}_{\beta,2}(a,b,k_1,k_2).
\label{intjpdf}
\end{equation}
The integration can be done in several steps, using that some of the integrals factorise.
We start with the integration over $b$ and choose polar coordinates.  Using \cite[3.382(4)]{Gradshteyn}, we obtain 
\begin{equation}
\int_{\mathbb{C}}{\rm d}^2b\ 
\frac{\exp[-2|b|^2]}{(4|b|^2+k_2^2)^\eta}
=\frac{\pi}{2^{1+\eta}}\exp\left[\frac{k_2^2}{2}\right]\Gamma\left[1-\eta,\frac{k_2^2}{2}\right],
\end{equation}
where $\Gamma[x,y]=\int_y^\infty t^{x-1}e^{-t}\mathrm{d}t$ is the incomplete gamma function. The integrations over $a=a_1+ia_2$ are simple shifted Gaussian integrals, leading to 
\begin{eqnarray}
&&\int_{-\infty}^{+\infty}\mathrm{d}a_1 \int_{-\infty}^{+\infty}\mathrm{d}a_2
e^{-2(a_1^2+a_2^2+k_2(a_1\cos k_1+a_2\sin k_1))}
\nonumber\\
&&=\frac{\pi}{2}\exp\left[
\frac{k_2^2}{2}(\cos^2(k_1)+\sin^2(k_1))
\right].
\end{eqnarray}
Consequently, the $k_1$-dependence drops out and integrating over $k_1$ yields $2\pi$. We are left with the integral over $k_2$ where the three exponentials cancel, leading to 
\begin{equation}
C_\beta^{-1}=\frac{\pi^3}{2^{1+\eta}}\int_0^\infty \mathrm{d}k_2 \Gamma\left[1-\eta,\frac{k_2^2}{2}\right]=\frac{\pi^3\sqrt{2}}{2^{1+\eta}} \Gamma\left[\frac{3}{2}-\eta\right].
\end{equation}
This step follows from \cite[6.455(1)]{Gradshteyn} and we arrive at the result claimed in \eqref{betaRMN2}.

For the normalisation $K_\beta$ of the joint probability density of complex eigenvalues in \eqref{betaZN2}, we have to show 
\begin{align}
	1=K_\beta \int_\mathbb{C}\mathrm{d}^2z_1 \int_\mathbb{C}\mathrm{d}^2z_2\  e^{-(|z_1|^2+|z_2|^2)}|z_2-z_1|^{2-2\eta}.
\end{align}
We change variables $(z_1,z_2)\rightarrow (z_1,z)$ with $z=z_2-z_1$, and choose polar coordinates as $z_1=r_1e^{i\Theta_1}$ and $z=r_2e^{i\Theta_2}$.
After this transformation, we find
\begin{equation}
\label{Kstep1}
		K_\beta^{-1}= 
\int_\mathbb{C}\mathrm{d}^2z_1 \int_\mathbb{C}\mathrm{d}^2z_2		
r_2^{2-2\eta}\ e^{-2r_1^2-r_2^2-2r_1r_2\cos(\Theta_1-\Theta_2)}.
\end{equation}
After using an addition theorem for cosine, one of the angular integration can be done, 
employing \cite[3.338(4)]{Gradshteyn}:
\begin{eqnarray}
&&\int_0^{2\pi} \mathrm{d}\Theta_1 \exp[b \sin \Theta_1 +c \cos \Theta_1]=2\pi I_0(y), \nonumber\\
&& y = \sqrt{b^2+c^2}=2r_1r_2.
\end{eqnarray}
Here $I_0(y)$ denotes the modified Bessel function of the first kind. The second angular integral then becomes trivial, giving $2\pi$, and we have
\begin{equation}
		K_\beta^{-1}=(2\pi)^2
\int_0^\infty \mathrm{d}r_1r_1 \int_0^\infty \mathrm{d}r_2 r_2^{3-2\eta}\   e^{-2r_1^2-r_2^2} I_0(2r_1r_2).
\end{equation}
The integral over $r_1$ follows, using \cite[6.631(4)]{Gradshteyn}:
\begin{equation}
\int_0^\infty \mathrm{d}r_1r_1\exp[-2r_1^2]\ I_0(2r_1r_2)=\frac{1}{4}\ e^{r_2^2/2}.
\end{equation}
For the remaining integral over $r_2$ we thus obtain
\begin{equation}
	K_\beta^{-1}=\pi^2\int_0^\infty \mathrm{d}r_2 r_2^{3-2\eta}\  e^{-r_2^2/2}=\pi^22^{1-\eta}\Gamma(2-\eta), 
\end{equation}
as was claimed in \eqref{betaZN2}, after inserting $\eta=1-\beta/2$.

\subsection{Spacing distribution without assuming translational invariance}\label{space}

Let us (re)derive the nearest neighbour spacing distribution in radial distance, 
$p_{\beta,2}(s)$, Eq. \eqref{peta} in the main text. There it was argued, that using translational invariance we may put one eigenvalue at the origin, $z_1=0$, and then compute the probability that the second eigenvalue $z_2$ is located at radius $s$. Here, we show that the same result holds when using the proper definition of the spacing distribution for $N=2$, 
\begin{align}
	p_{\beta,2}(s)=\int_\mathbb{C} \mathrm{d}^2z_1 \int_\mathbb{C} \mathrm{d}^2z_2 \mathcal{P}_{\beta,2}(z_1,z_2)\delta(s-|z_2-z_1|).
\end{align}
It describes the probability to find the two eigenvalues at distance $s$ from another. Because of $N=2$, there are no further eigenvalues present. Using the very same change of variables from the previous subsection before \eqref{Kstep1}, we obtain
\begin{eqnarray}
p_{\beta,2}(s) &=&K_\beta  
\int_0^{2\pi} \mathrm{d}\Theta_1 \int_0^\infty \mathrm{d}r_1r_1\int_0^{2\pi} \mathrm{d}\Theta_2\int_0^\infty \mathrm{d}r_2r_2
\nonumber\\
&&\times r_2^{2-2\eta}  e^{-(2r_1^2+r_2^2+2r_1r_2\cos(\Theta_2-\Theta_1))} \delta(s-r_2)
\nonumber\\
&=&K_\beta \pi^2s^{3-2\eta}e^{-s^2/2}.
\end{eqnarray}
Because we can follow the same steps of integration we just give the final answer here. It is easy to check that it is indeed properly normalised. However, we still need to rescale the first moment to unity, which reads
\begin{equation}
m=\int_0^\infty ds\, s\, p_{\beta,2}(s)=2^{\frac12}\frac{\Gamma\left(\frac{5}{2}-\eta\right)}{\Gamma(2-\eta)}.
\end{equation}
Thus our final answer for the spacing distribution with normalised first moment $\hat{p}_{\beta,2}(s)$ is
\begin{eqnarray}
\hat{p}_{\beta,2}(s)=mp_{\beta,2}(ms)&=& \frac{2\alpha^{2-\eta}}{\Gamma(2-\eta)}\ s^{3-2\eta}\exp[-\alpha s^2], \nonumber\\
\alpha&=&\frac{\Gamma\left(\frac{5}{2}-\eta\right)^2}{\Gamma(2-\eta)^2}.
\end{eqnarray}
It agrees with the quantity given in \eqref{peta} in the main text, after dropping the hat $\hat{}$ and inserting $\beta=2-2\eta$.\\

\section{Comparison of Polynomials Fits for $\beta_{\rm eff}$
}\label{fits}

In this appendix we provide some further details about the determination of the fitting function $\beta_{\rm eff}(\beta)$ that is used in our surmise \eqref{cubic} in Section \ref{Surmise}. In particular, we compare the quality of different approximations for $\beta_{\rm eff}(\beta)$, using a linear, quadratic or cubic fit and give their standard deviations and Kolmogorov-Smirnov distances to the best fit $\beta_{\rm eff}$. 

\begin{figure}[h!]
		\centering
		\includegraphics[width=0.79\linewidth,angle=0]{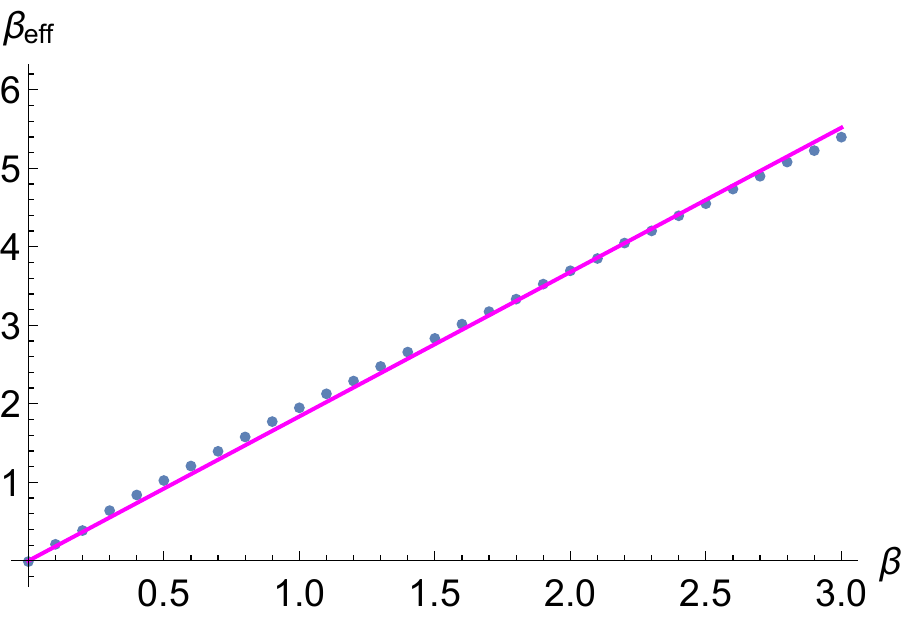}
		\caption{Linear fit of $\beta_{\rm eff}(\beta)$ \eqref{eq-beta degree1} (full line) to the best fitted value of $\beta_{\rm eff}$ in \eqref{petasurmise} (points).}
		\label{Fig:betafit-linear}
	\end{figure}

In Fig. \ref{Fig:betafit-linear} we first of all show a linear fit \eqref{eq-beta degree1}  to the data points that were obtained by optimising the value of $\beta_{\rm eff}$ in \eqref{petasurmise} in comparison to the numerically generated spacing of the 2DCG, in steps of $0.1$.

While to first approximation the points follow a straight line, there are visible deviations below and above $\beta=2$. We have thus improved the fit by using a quadratic, respectively cubic polynomial, resulting into
	\begin{align}
		\beta_{\text{eff},1}&=1.839\beta, \label{eq-beta degree1}\\
		\beta_{\text{eff},2}&=1.999\beta-0.070\beta^2, \label{eq-beta degree2}\\
		\beta_{\text{eff},3}&=2.108\beta-0.190\beta^2+0.030\beta^3. \label{eq-beta degree3}
	\end{align}
The corresponding cubic fit \eqref{eq-beta degree3} is shown in Section \ref{Surmise} in Fig. \ref{Fig:betafit-cubic}, where deviations are no longer visible by eye.

In Table \ref{tab:sigmas} we compare the standard deviation \eqref{sigma-def} of the linear and cubic fit to the  best fit. 
Already the linear fit is almost indistinguishable form the best fit from $\beta\geq 1.2$, whereas the cubic fit only deviates from the best fit in the first three entries with $\beta=0.1-0.3$. 
For that reason we do not display the corresponding values for the quadratic fit \eqref{eq-beta degree2}.
The same picture emerges when comparing the Kolmogorov-Smirnov distances \eqref{KS-def} which are independent of the binning. They are shown in Table \ref{tab:KS}.

	\begin{table}[h]\centering
		\caption[]{List of standard deviations \eqref{sigma-def} 
in units $ 10^{-2}$ using a linear fit $\sigma_1$ \eqref{eq-beta degree1}, cubic fit $\sigma_3$ \eqref{eq-beta degree3}, and best fit $\sigma_b$ (points in Fig. \ref{Fig:betafit-linear}).}\vspace{3mm}
		\begin{tabular}{|c|c|c|c|c|c|c|c|c|c|c|c|c|c|c|c|}
			\hline
			$\beta$  & 0.1  & 0.2 & 0.3 & 0.4 & 0.5 & 0.6 & 0.7 & 0.8& 0.9  &1.0 &1.1&1.2&1.3&1.4&1.5\\
			\hline\hline
			$\sigma_1$ & 0.9  & 1.6 & 2.0 & 2.4 & 2.7 & 2.9 & 3.2 & 3.5 &3.6  &3.8 & 3.9 & 4.0  & 4.1 & 4.2 & 4.3\\
			\hline
			$\sigma_3$ & 0.9  & 1.4 & 1.9 & 2.3 & 2.6 & 2.8 & 3.1 & 3.4 &3.6  &3.7 & 3.8 & 4.0  & 4.1 & 4.2 & 4.3\\
			\hline		
						$\sigma_b$ & 0.8  & 1.5 & 1.9 & 2.2 & 2.6 & 2.8 & 3.1 & 3.4 &3.6  &3.7 & 3.8 & 4.0  & 4.1 & 4.2 & 4.3\\
			\hline				
			\hline
			$\beta$ & 1.6&1.7&1.8&1.9&2.0& 2.1 & 2.2  & 2.3 & 2.4 & 2.5 & 2.6 & 2.7 & 2.8 & 2.9& 3.0  \\
			\hline\hline
			$\sigma_1$ & 4.4 & 4.5 &4.5 & 4.6& 4.7 & 4.9 & 4.9  & 4.9 & 4.9 & 5.2 & 5.3& 5.2 &5.2 & 5.4& 5.4   \\\hline
						$\sigma_3$ & 4.3 & 4.5 &4.5 & 4.6& 4.7 & 4.9 & 4.9  & 4.9 & 4.9 & 5.2 & 5.3& 5.2 &5.2 & 5.4& 5.4   \\\hline
						$\sigma_b$ & 4.3 & 4.5 &4.5 & 4.6& 4.7 & 4.9 & 4.9  & 4.9 & 4.9 & 5.2 & 5.3& 5.2 &5.2 & 5.4& 5.4   \\\hline
		\end{tabular}
		\label{tab:sigmas}
	\end{table}
	

	\begin{table}[h]\centering
		\caption[]{List of Kolmogorov-Smirnov distances \eqref{KS-def} in units $ 10^{-2}$ 
		using a linear fit $d_1$ \eqref{eq-beta degree1}, cubic fit $d_3$ \eqref{eq-beta degree3}, and best fit $d_b$ (points in Fig. \ref{Fig:betafit-linear}).}\vspace{3mm}
		\begin{tabular}{|c|c|c|c|c|c|c|c|c|c|c|c|c|c|c|c|}
			\hline
			$\beta$  & 0.1  & 0.2 & 0.3 & 0.4 & 0.5 & 0.6 & 0.7 & 0.8& 0.9  &1.0 &1.1&1.2&1.3&1.4&1.5\\
			\hline\hline
					$d_1$ & 0.8  & 0.8 & 1.1 & 1.3 & 1.5 & 1.6 & 1.7 & 1.8 &1.9  &2.0&2.1 & 2.1  & 2.1 & 2.1 & 2.2 \\ \hline
			$d_{3}$ & 0.7  & 0.6 & 0.9 & 1.1 & 1.3 & 1.5 & 1.6 & 1.7 &1.8  &1.9&2.0 & 2.0  & 2.1 & 2.1 & 2.1 \\ \hline
						$d_{b}$ & 0.7  & 0.8 & 0.8 & 1.1 & 1.3 & 1.5 & 1.6 & 1.7 &1.8  &1.9&2.0 & 2.0  & 2.1 & 2.1 & 2.2 \\
			\hline
			\hline
			$\beta$ & 1.6&1.7&1.8&1.9&2.0& 2.1 & 2.2  & 2.3 & 2.4 & 2.5 & 2.6 & 2.7 & 2.8 & 2.9& 3.0  \\
			\hline\hline
				$d_1$ &  2.2 & 2.3 &2.3 & 2.3& 2.3 & 2.4 & 2.3  & 2.3 & 2.4 & 2.4 & 2.4& 2.4 &2.3 & 2.4& 2.4    \\ \hline
			$d_{3}$ &  2.2 & 2.2 &2.2 & 2.3& 2.3 & 2.4 & 2.3  & 2.3 & 2.4 & 2.4 & 2.4& 2.4 &2.4 & 2.4& 2.4    \\ \hline
					$d_{b}$ &  2.2 & 2.2 &2.2 & 2.3& 2.3 & 2.4 & 2.3  & 2.3 & 2.4 & 2.4 & 2.4& 2.4 &2.4 & 2.4& 2.4    \\
			\hline
		\end{tabular}
		\label{tab:KS}
	\end{table}

\end{appendix}


\begin{thebibliography}{99}
\bibitem{nounfold} As in the derivation for the Wigner surmise \eqref{pWigner} in 1D, we do not unfold here in 2D either. This is because at $N=2$ there is no separation between bulk and edge. Fixing normalisation and first moment to unity makes these spacing distributions in 1D and 2D unique.
\end{thebibliography}
\end{document}